\documentclass[10pt, conference]{IEEEtran}

\usepackage[margin = 0.7in]{geometry}
\usepackage{amsmath}
\usepackage{algpseudocode}
\usepackage{amssymb}
\usepackage{amsfonts}
\usepackage{amscd}
\usepackage{mathrsfs}
\usepackage{graphicx}
\usepackage[usenames,dvipsnames]{color}
\usepackage{bm}
\usepackage{url}
\usepackage{verbatim}
\usepackage[mathcal]{euscript}

\newtheorem{theorem}{Theorem}

\newtheorem{Proposition}{Proposition}

\newtheorem{remark}{Remark}

\begin{document}

\title{Convolution Attack on Frequency-Hopping by Full-Duplex Radios}

\author{J. Harshan$^{\dagger}$ and Yih-Chun Hu$^{\ast}$\\
$^{\dagger}$Indian Institute of Technology Delhi, India, $^{\ast}$University of Illinois Urbana-Champaign, USA\\
Email: jharshan@ee.iitd.ac.in, yihchun@illinois.edu\\}


%

\maketitle

\begin{abstract}
We propose a new adversarial attack on frequency-hopping based wireless communication between two users, namely Alice and Bob. In this attack, the adversary, referred to as Eve, instantaneously modifies the transmitted signal by Alice before forwarding it to Bob within the symbol-period. We show that this attack forces Bob to incorporate Eve's signal in the decoding process; otherwise, treating it as noise would further degrade the performance akin to jamming. Through this attack, we show that Eve can convert a slow-fading channel between Alice and Bob to a rapid-fading one by modifying every transmitted symbol independently. As a result, neither pilot-assisted coherent detection techniques nor blind-detection methods are directly applicable as countermeasures. As potential mitigation strategies, we explore the applicability of frequency-hopping along with (i) On-Off keying (OOK) and (ii) Binary Frequency-Shift-Keying (FSK) as modulation schemes. In the case of OOK, the attacker attempts to introduce deep-fades on the tone carrying the information bit, whereas in the case of BFSK, the attacker pours comparable energy levels on the tones carrying bit-$0$ and bit-$1$, thereby degrading the performance. Based on extensive analyses and experimental results, we show that (i) when using OOK, Bob must be equipped with a large number of receive antennas to reliably detect Alice's signal, and (ii) when using BFSK, Alice and Bob must agree upon a secret-key to randomize the location of the tones carrying the bits, in addition to randomizing the carrier-frequency of communication. \end{abstract}

\begin{IEEEkeywords}
Jamming, frequency-hopping, cognitive radio, convolution attack, wireless security
\end{IEEEkeywords}
%
\section{Introduction}

Jamming is a well known adversarial attack on wireless communication \cite{Yulong}, \cite{J1}, \cite{J8}, \cite{J9}, \cite{Jamm1} wherein the attacker overpowers the communication between a transmitter and a receiver by injecting high-powered noise signals. Standard ways to mitigate jamming include frequency-hopping (FH) \cite{PKK1}, \cite{PKK2} and direct sequence spread spectrum (DSSS) schemes \cite{proakis}. In the case of DSSS, a narrowband signal is spread across a wide band of frequencies by using a spreading code so that an attacker, which does not possess the spreading code, will have its jamming signal rejected by the receiver. In the case of FH, which is the subject matter of this paper, the transmitter and the receiver synchronously hop across several carrier-frequencies so that the hopping pattern appears non-deterministic to the adversary. As a result, narrowband jamming, i.e., jamming a specific carrier-frequency, cannot guarantee performance degradation due to randomness in the hopping pattern. On the other hand, with wideband jamming, i.e., jamming all the carrier-frequencies, the effective noise power injected on each carrier-frequency would be too weak to induce significant degradation in the performance. Both DSSS and FH are effective under the assumption that the attacker is power-constrained. Citing these benefits, DSSS and FH have found extensive applications in military communication systems, and recently in many cyber-physical systems. While DSSS and FH introduce randomness in the choice of the spreading code and carrier-frequencies, respectively, introducing randomness over spatial orientation of the antennas has also been explored as a viable anti-jamming technique in communication involving highly-directional antennas \cite{Jerry_Hu}. 


With wireless communication being an integral part of most cyber-physical systems, e.g. urban transportation, smart-grid and other IOT systems \cite{Jamm2}, it is imperative to envision new attacks \cite{HaY} on such systems and provide suitable countermeasures against them. Over the past decade, wireless communication technology has witnessed enormous progress in bandwidth-efficient physical-layer techniques that have helped wireless devices achieve high data-rate. One of the prominent areas rising in this space is full-duplex communication \cite{DuSa}, \cite{CJSLK}, \cite{mayank_jain}, wherein a radio device can simultaneously transmit and receive signals in the same frequency band. While efficient hardware- and software-architectures have helped full-duplex radios to achieve near-perfect self-interference cancellation, there have been concurrent developments in hardware implementation for low-latency processing of radio frequency (RF) signals \cite{RaCa} in the field of systems security. Aggregating the latest developments in the above areas, we believe that next-generation cyber-physical systems ought to assume strong attack models that employ state-of-the-art wireless techniques.

\subsection{Motivation}

\begin{figure}
\begin{center}
\includegraphics[width=9cm]{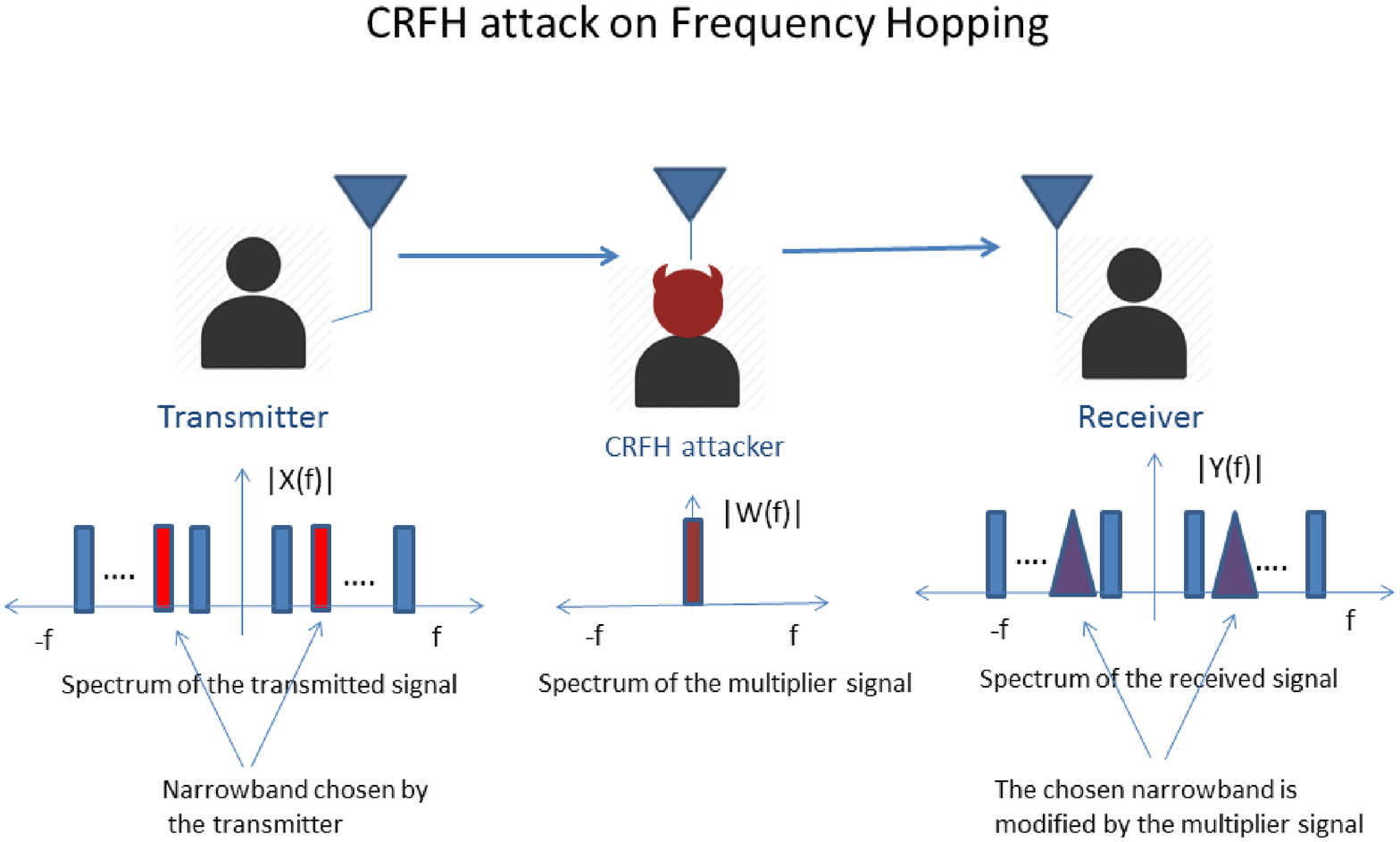}
\caption{\label{fig:CRFH_FH_conv_idea}Under the framework of Cognitive Radio from Hell (CRFH), the attacker can modify the signal in any narrowband chosen by the two users despite not knowing the hopping pattern. In the above figure, $X(f)$, $W(f)$, and $Y(f)$, respectively, denote the Fourier transform of the transmitted passband signal, attacker's random signal, and the received passband signal.}
\end{center}
\end{figure}

In this paper, we are interested in threat models arising out of full-duplex radios that operate as hidden relays between a transmitter and a receiver, as depicted in Fig. \ref{fig:CRFH_FH_conv_idea}. Loosely speaking, this threat comes under the well known framework of \emph{man-in-the-middle attacks}, wherein the attacker can manipulate the transmitted symbols before they reach the legitimate receiver. Although instantaneous modification of transmitted symbols has been addressed to mitigate interference in wireless networks \cite{HJo, WDZLQL, Uppal}, such ideas have not been studied as a threat to wireless security. An important question to answer along that direction is: \emph{Are the current-day wireless systems resilient to instantaneous manipulation of transmitted symbols in the air?} From the standpoint of practicality, major challenges to instantaneous modification include (i) additional processing-delay and (ii) additional path-delay, introduced by the attacker. The processing-delay constraint restricts the attacker not to make dynamic decisions based on baseband signal processing of the received signals. On the other hand, the path-delay constraint resulting from the attacker's position restricts it to be appropriately placed so that the forwarded components reach the receiver well within the delay of one symbol-period relative to the signals received from the main path. In a nutshell, if the above two constraints are respected, then, in principle, it is possible for the modified signals to arrive at the legitimate receiver within the symbol-period. We refer to such an attacker as Cognitive Radio from Hell (CRFH). The proposed adversarial model comes under the class of correlated jamming \cite{CJ2} wherein the jammer has full or partial information about the transmitter's signals.
	
\subsection{CRFH Attacks on Frequency-Hopping}

Our discussion in the preceding section indicates that the processing-delay and the path-delay constraints may preclude the attacker from executing instantaneous modification on wideband communication due to small symbol-periods. However, on narrowband communication systems, such as FH (wherein the bandwidth around the chosen carrier-frequency is small), the attacker can potentially execute instantaneous modification ``in the air" due to relatively large symbol-periods compared to the path-delay on the main path. This has motivated us to study the effect of CRFH attacks on FH systems. Particularly, in the context of FH, we note that instantaneous modification of symbols is crucial to degrade the error-performance at Bob, otherwise, any unintentional delay introduced by the attacker, will result in a delay of one symbol-period or more. This allows the legitimate receiver to evade the attack by hopping to the next carrier-frequency before the delayed components arrive. Thus, to enforce degraded performance from instantaneous modification, the CRFH attacker on FH must respect the delay constraints on instantaneous modification.

\subsection{Related Work}
\label{sec:related_works}

To help translate the idea of CRFH attack to practice, recent advances in the field of full-duplex radios \cite{DuSa, CJSLK, mayank_jain, Zhou_eta_al, ANOGT, AARRU} have shown that radios can be designed to cancel their self-interference while instantaneously forwarding the transmitted signals. In particular, \cite{DinKa_fast} has showcased the possibility of building radios with the capability of instantaneous processing and relaying. Recently, \cite{Uppal} has demonstrated the effectiveness of instantaneous modification by full-duplex radios to achieve co-existence in interference channels. Also, \cite{RaCa} has showed that radio-frequency signals can be processed and retransmitted in the analogue domain with a delay of few nano-seconds.

The notion of modifying symbols in the air is also known under the framework of reactive jamming  \cite{J7}, which refers to the process of targeting selected packets in the air as it allows the attacker to destroy specific packets and yet go undetected. The authors of \cite{J2} have studied the feasibility of reactive jamming by designing and implementing a reactive jammer against 802.15.4 networks. Through the use of Universal Software Radio Peripherals (USRPs), \cite{J2} has demonstrated jamming attack with reaction time of the order of microseconds in indoor environments. In \cite{J4}, the authors have addressed reactive jamming attacks where the adversary is capable of picking packets based on real-time classification of packets at the physical-layer. They have also proposed countermeasures to prevent real-time packet classification based on both cryptographic as well as physical-layer ideas. In \cite{J5}, the authors have proposed a technique to detect reactive jammers on DSSS. The basic idea is to use statistics on attack-free packets and then identify packets attacked from those lost due to bad channel condition. In \cite{J6}, reactive jamming on Orthogonal Frequency Division Multiplexing (OFDM) is considered, and an effective countermeasure based on Multiple-Input Multiple-Output (MIMO) systems has been proposed. Overall, inspired by the above works, particularly that of \cite{DinKa_fast} and \cite{RaCa}, we believe that it is imperative for existing cyber-physical systems to envision attacks that could arise out of full-duplex radios capable of instantaneous modification of transmitted symbols.


\subsection{Contributions}

The contributions of this paper are summarized below:

\begin{itemize}
\item We introduce a new adversarial attack, referred to as the convolution attack (CA), on FH based wireless communication. In this attack, the adversary, which is strategically positioned between the transmitter and the receiver, instantaneously multiplies the received passband signal by a random baseband signal, and then forwards it to the receiver within the symbol-period. Subsequently, the forwarded signals will combine with the signals directly received from the transmitter, thereby modifying the information symbols \emph{in the air}. One of the highlights of the attack is that the attacker is able to perturb the transmitted symbols despite not knowing the active narrowband of communication. We show that the proposed attack forces the legitimate receiver to incorporate the forwarded signals in the decoding process; otherwise, discarding them as noise is shown to result in consequences akin to jamming. (see Section \ref{sec:chall_CRFH}). We also show that the CA forces the equivalent channel between the legitimate users to experience frequency-selectivity and rapid-fading, in such a way that neither pilot-based coherent detection techniques, nor traditional non-coherent and differential encoding/decoding detection techniques can mitigate the attack. 

\item As a countermeasure against CA, we study the performance of an FH system with non-coherent On-Off Keying (OOK) as the underlying modulation scheme (see Section \ref{sec:on_off}). This mitigation strategy, although a traditional communication scheme, is tailor-made to handle the threat model because switching-off the transmission forbids the attacker from perturbing the communication, while switching-on the transmission helps the receiver to collect energy despite the attack. With large number of antennas at the receiver, we show that the receiver can opportunistically use the attacker's signal to its advantage to gather more energy for detection (see Section \ref{subsec:cross_1_0}). 
We show that OOK is an effective countermeasure if the attacker executes the convolution attack on both the pilot symbols and the data symbols with the same attack parameters; this is because the threshold for energy detection is computed using the received energy distributions on the pilot symbols. However, if the attacker decides to selectively attack only the data symbols and not the pilots, then OOK is no longer an effective countermeasure.

\item As a second countermeasure against CA, we study the performance of non-coherent Binary Frequency Shift Keying (BFSK), a widely used modulation scheme with FH in military application. In this form of CA, the adversary instantaneously modifies the signal ``in the air" so that the receiver witnesses comparable energy levels on the tones carrying bit-$0$ and bit-$1$. As a result, this attack significantly degrades the error performance at the receiver when it uses threshold-based energy detection (see Section \ref{sec:CA_on_FSK}). One the fundamental causes for this attack is that although the carrier-frequency is randomly hopped based on a shared secret-key between the legitimate users, the locations of the tones carrying bit-$0$ and bit-$1$ are deterministic upto one bit randomness when the attacker observes the signal in the air. We first show that this form of CA, when appropriately executed, introduces error-floor behaviour in the error performance. Subsequently, we propose a mitigation strategy, referred to as Enhanced BFSK, wherein unlike the standard frequency-hopping technique, the tones carrying bit-$1$ and bit-$0$ are also randomized based on an additional secret-key shared between the legitimate users. As a consequence, upon observing the transmitted signal in the air, the attacker continues to have uncertainty about the location of tone carrying the complementary bit, thereby forcing it to execute wideband jamming. Unlike the case of OOK, we show that BFSK is resilient even if the attacker selectively executes the CA on the data symbols and not on the pilots; this is because BFSK detection does not rely on the distribution of received energy on the pilots.
\end{itemize}

Henceforth, throughout the paper, we refer to the transmitter, the receiver, and the attacker as Alice, Bob and Eve, respectively. To model a power-constrained attacker, we assume that Eve has $\theta$ times more energy than Alice, i.e., $E_{Eve} = \theta E_{Alice}$, for some $\theta >> 1$. This implies that as Alice increases her energy to improve the performance, Eve can also proportionately increase her energy. Furthermore, out of $E_{Eve}$, Eve may use only a fraction of it, say $E_{Eve, C} = \alpha E_{Eve}$, for some $0 \leq \alpha \leq 1$, on the CA. Thus, the key attack parameters of this paper are $\theta >> 1$ and $0 \leq \alpha \leq 1$.


\section{Convolution Attack on Frequency-Hopping}
\label{sec:CRFH_FH}

Consider an FH based amplitude-modulated communication scheme between Alice and Bob, wherein the carrier-frequency $f_{c}$ of the transmitted narrowband signal is randomly chosen from one of the $N$ tones, denoted by the set $\mathcal{F} = \{f_1, f_2, \ldots, f_N\}$. Let Alice use a root raised cosine (RRC) waveform \cite{proakis}, denoted by $g(t)$, as the baseband signal of bandwidth $W$ Hz and symbol rate $T$ seconds. Furthermore, let $\{x_{k} = x_{I, k} + \imath x_{Q, k} ~|~ k = 0, 1, 2, \ldots\}$, with $\imath = \sqrt{-1}$, denote the sequence of complex symbols, where $x_{k}$ takes value from a $2$-dimensional finite complex constellation, e.g. quadrature amplitude modulation. The corresponding train of baseband signals is given by
\begin{equation}
\label{eq:baseband1}
s_{b}(t)  =  \sum_{k} \sqrt{E_{Alice}} x_{k}g(t - kT),
\end{equation}
where $E_{Alice}$ is the average transmit energy by Alice assuming that $x_{k}$ and $g(t)$ are appropriately normalized. After modulating $s_{b}(t)$ on carrier-frequency $f_{c} \in \mathcal{F}$, the transmitted passband signal is of the form
\begin{eqnarray}
\label{eq:basebandeq}
s(t) = \mathcal{R}\left(s_{b}(t)e^{2 \pi \imath f_{c}t}\right).
\end{eqnarray}
where $\mathcal{R}(\cdot)$ denotes the real part of a complex number. The set $\mathcal{F}$ is chosen such that $|f_{i} - f_{i+1}| > W$, for $1 \leq i \leq N-1$, thereby leaving sufficient guard-band to mitigate inter-carrier interference. Alice employs a carrier-frequency $f_{c} \in \mathcal{F}$ for $T_{hop} = mT$ seconds, for some integer $m > 0$, before hopping to another value in $\mathcal{F}$. Meanwhile, Bob synchronously hops across the same sequence of carrier-frequencies every $T_{hop}$ seconds, as the hopping pattern is generated using a shared secret key. Throughout the paper, we use small values of $m$ by assuming that Alice and Bob are capable of quickly switching the carrier-frequencies with minimal losses due to transients in the transmit and receive RRC filters. Otherwise, with $m >> 1$, a sophisticated attacker can sense the narrowband of communication and subsequently inject jamming energy on the detected band, akin to traditional jamming. Thus, small values of $m$ helps the legitimate users to mitigate standard jamming attacks on frequency-hopping. The discrete-time version of the received baseband signal at Bob is given by
\begin{equation}
\label{eq:no_attack_channel}
y_{k}(f_{c}) = \sqrt{E_{Alice}} h^{(AB)}_{k}(f_{c})x_{k} + n^{(B)}_{k}(f_{c}), 
\end{equation}
for $k = 0, 1, \ldots, m -1$, where $h^{(AB)}_{k}(f_{c})$ is the complex channel gain on the tone $f_{c}$, and $n^{(B)}_{k}(f_{c})$ is the additive noise at Bob, distributed as $\mathcal{CN}(0, \sigma^{2}_{Bob})$. A complex random variable $g \sim \mathcal{CN}(0, \sigma^{2})$ is said to be circularly symmetric Gaussian distributed when the real and imaginary components of $g$ are Gaussian and i.i.d. with mean $0$ and variance $\frac{\sigma^{2}}{2}$. We assume that the channel $h^{(AB)}_{k}(f_{c})$ remains fixed within the hopping interval, i.e., $h^{(AB)}_{k}(f_{c}) = h^{(AB)}(f_{c}), 0 \leq k \leq m-1$. For brevity, henceforth, we drop the reference to the carrier-frequency $f_{c}$ from the channel model in \eqref{eq:no_attack_channel}.

We assume that Eve is positioned somewhere between Alice and Bob, and she is not aware of the secret key used to generate the hopping pattern. At any point in time, due to the non-deterministic hopping pattern, Eve cannot successfully guess the \emph{active} carrier-frequency with probability one, and therefore narrowband jamming on a random carrier-frequency in $\mathcal{F}$ does not guarantee degraded error performance.

Keeping in view practical hurdles in executing wideband jamming by a power-constrained attacker, we envision a new attack, referred to as the \emph{convolution attack}, which is as depicted in Fig. \ref{fig:CRFH_FH_conv_idea}. We assume that Eve is capable of receiving and transmitting signals in the entire wideband covering all the $N$ carrier-frequencies. Through the CA, we show that Eve can modify the transmitted narrowband signal despite not knowing the active carrier-frequency. We draw our inspiration from the fact that analogue processing of narrowband signals is feasible in negligible amount of time \cite{RaCa}. In the proposed attack, Eve multiples the received passband signal with a random baseband signal, denoted by $w(t)$, in the analogue domain, and then forwards it to Bob. This multiplication operation in the time-domain is equivalent to convolving the Fourier transform of the received signal with that of the random signal. Because of this operation, Eve can modify the narrowband signal without knowing the carrier-frequency. Subsequently, this modified version of the signal is added to the signal arriving directly from Alice, thus corrupting the overall received signal in the narrowband of interest. We make the following assumptions for executing the CA: (i) Negligible processing-delay at Eve, (ii) Negligible path-delay through Eve, and (iii) Full-duplex architecture with perfect cancellation at Eve.

In the next subsection, we mathematically describe how the forwarded signals from Eve affect the received signal at Bob.

\subsection{Signal Model with Convolution Attack}

Since Eve does not know the active carrier-frequency, she receives signals in the entire band, covering all the $N$ carrier-frequencies. Specifically, the received signal is given by
\begin{equation*}
r(t) = \sum_{i} a^{(AE)}_{i} s(t - \tau^{(AE)}_{i}) + z(t),
\end{equation*}
where $i$ denotes the $i$-th multipath component from Alice to Eve, and $a^{(AE)}_{i}$ and $\tau^{(AE)}_i$, respectively denote the corresponding amplitude and delay associated with the multipath component. Once Alice and Bob are locked onto a carrier-frequency, we assume that the rest of the $N - 1$ narrowbands are unused by other users in the network, and therefore, the non-signal component of $r(t)$ constitutes only the additive noise at Eve. In general, when the $N$ narrowbands are shared among several users in the network, the received signal $r(t)$ also constitutes interference from other users. Although Eve receives over a wide band of frequencies, we assume that the channel from Alice to Eve is frequency-flat over the active narrowband. 
Upon receiving $r(t)$, Eve multiplies it by a real random signal $w(t)$ (of unit average-energy over the symbol-period), and then transmits
\begin{equation}
\label{eq:multiplier}
e(t) = \sqrt{\alpha \theta }r(t)w(t),
\end{equation}
where $\sqrt{\alpha \theta}$ is the gain introduced by Eve for some $\theta >> 1$ and $0 \leq \alpha \leq 1$. The product operation $r(t)w(t)$ can be viewed as a way of introducing Doppler shifts to the passband signal by various frequency components of $w(t)$. With $e(t)$ transmitted from Eve, the received signal at Bob is given by
\begin{eqnarray}
\label{eq:rx_signal_bob}
y(t) & = & \sum_{j} b^{(EB)}_{j} e(t - \tau^{(EB)}_{j} - t_{p})\nonumber \\
& & + \sum_{q} c^{(AB)}_{q} s(t - \tau^{(AB)}_{q}) + n(t),
\end{eqnarray}
where the first part is contributed by Eve, the second part comes directly from Alice, and the last part $n(t)$ is the ambient noise generated at Bob's receiver. In \eqref{eq:rx_signal_bob}, $b^{(EB)}_{j}$ and $\tau^{(EB)}_j$, respectively denote the amplitude and the delay associated with the $j$-th multipath component from Eve to Bob. Similarly, $c^{(AB)}_{q}$ and $\tau^{(AB)}_q$, respectively denote the amplitude and the delay associated with the $q$-th multipath component from Alice to Bob.
Also, note that $t_{p}$ is the processing-delay introduced by Eve when multiplying the two signals. Among the multipath components from Alice to Bob, let $\tau^{AB}_{f}$ denote the first significant multipath component. Similarly, among the multipath components from Alice to Eve, and Eve to Bob, let $\tau^{AE}_{f}$ and $\tau^{EB}_{f}$ denote the first significant multipath components, respectively. In the proposed attack, Eve positions herself such that the following condition on delay is satisfied:
\begin{equation}
\label{eq:delay_constraint}
\tau^{(AB)}_{f} < \tau^{(AE)}_{f} + t_{p} + \tau^{(EB)}_{f} < \tau^{(AB)}_{f} + T.
\end{equation}
If the timing constraint in \eqref{eq:delay_constraint} is satisfied, then it is straightforward to verify that Eve's signal $w(t)$ can modify the current symbol in the air. After downconverting the received signal $y(t)$ from the carrier-frequency $f_{c} \in \mathcal{F}$, and then sampling and filtering, we obtain the discrete-time version of the baseband received signal, given by
\begin{eqnarray}
\label{eq:selective_model}
y_{k} =  \sqrt{E_{Alice}}h^{(AB)}_{k}x_{k} + \sum_{l = 0}^{L_{k}-1}  h^{(AEB)}_{k, l}\sqrt{\alpha \theta E_{Alice}}x_{k - l}\nonumber \\
+ \sqrt{\alpha \theta}n^{(EB)}_{k} + n^{(B)}_{k},
\end{eqnarray}
for $k = 0, 1, \ldots, m -1$, where $\{h^{(AEB)}_{k, l} ~|~ 1 \leq l \leq L_{k}\}$ are the complex channels contributed by Eve, $n^{(EB)}_{k}$ is the noise component forwarded by Eve, and $n^{(B)}_{k}$ is the additive noise at Bob. The channel contributed by Eve is possibly frequency-selective, where the number of taps of the channel, denoted by $L_{k}$, depends on the chosen waveform $w(t)$. Intuitively, as depicted in Fig. \ref{fig:CRFH_FH_conv_idea}, although the channel from Alice to Eve, and Eve to Bob are frequency-flat within a narrowband of $W$ Hz, the convolution operation in the frequency domain can disrupt the frequency-flat structure, thereby giving rise to a frequency-selective channel.

Observe that Eve is not injecting noise into the narrowband of interest, instead she is instantaneously modifying the transmitted symbols by a random quantity, which is some complex function of (i) the channel from Alice to Eve, (ii) the signal $w(t)$, and (iii) the channel from Eve to Bob. If the timing constraint in \eqref{eq:delay_constraint} is not satisfied, then the signal forwarded by the attacker does not modify the current symbol in the air, instead it reaches Bob in the subsequent symbol-periods. This implies that $h^{(AEB)}_{k, 0} = 0$ in \eqref{eq:selective_model}. Although this form of attack continues to affect the signal-to-noise-ratio of subsequent symbols, the current symbol in the air does not get modified. A straightforward way for Alice and Bob to evade this attack is by locking to a given carrier-frequency for just one symbol before hopping to another carrier-frequency in $\mathcal{F}$. Thus, satisfying the timing constraint in \eqref{eq:delay_constraint} is crucial for Eve to execute the CA when the legitimate users have the potential to hop carrier-frequencies with $m = 1$.

\section{Challenges in Mitigating Convolution Attack}
\label{sec:chall_CRFH}

Without the attack, i.e., $h^{(AEB)}_{k, l} = 0, \forall k, l$, the complex channel $h^{(AB)}_{k}$ is determined only by the environment. Importantly, the coherence-time of the channel $h^{(AB)}_{k}$ is determined only by the relative velocity of the surrounding objects in the environment. However, with attack, an additional signal component $\sum_{l = 0}^{L_{k}-1}  h^{(AEB)}_{k, l}x_{k - l}$ is added to the received signal at Bob as shown in \eqref{eq:selective_model}. A naive way to handle this additional term is by considering it as noise. However, this will naturally lower the signal-to-noise-ratio (SINR), and degrade the error performance when Eve's power is dominant. Instead, since the additional component contains useful information, it is prudent for Bob to treat it as the signal term in the decoding process. After incorporating Eve's signals in the decoding process, Bob is forced to view an equivalent channel model, given by
\begin{equation}
\label{eq:eq_signal_model_with_attack}
\sqrt{E_{Alice}}h^{(AB)}_{k}x_{k} + \sum_{l = 0}^{L_{k}-1}  h^{(AEB)}_{k, l}\sqrt{\alpha \theta E_{Alice}}x_{k - l}.
\end{equation}
Although Eve is contributing additional signal power into the system, Bob is unsure of how to use this additional power as it may rapidly change every symbol. We now summarize the major changes introduced in the channel model when Eve executes CA with significant power compared to that at Alice: (i) Since $w(t)$ can be arbitrarily chosen by Eve, the equivalent channel can be frequency-selective despite using narrowband for communication. (ii) Unlike in traditional channels, the delay-spread of the equivalent frequency-selective channel may change each symbol since $w(t)$ could be composed of arbitrary segment of signals every $T$ seconds, and finally, (iii) the coherence-time of the equivalent channel can also be controlled by Eve, to the extent that the channel seen across two successive symbols can be uncorrelated. It is worth emphasizing that Eve is able to force abrupt variations in two fundamental characteristics of the channel, namely: frequency-selectivity and Doppler-spread. To bring in these variations, it is necessary for Eve to spend significant power compared to Alice, otherwise the characteristics of the true wireless channel will continue to dominate, and as a result the attack will be ineffective. 

\begin{figure*}
\begin{small}
\begin{equation}
\label{eq:on-off-keying_rx_signal}
y_{k} = \left\{ \begin{array}{cccccccccc}
\sqrt{E_{Alice}}h^{(AB)}_{k} + \sqrt{\alpha\theta E_{Alice}}h^{(AEB)}_{k} + \sqrt{\alpha\theta}n^{(EB)}_{k} + n^{(B)}_{k}, & \mbox{ if } b_{k} = 1;\\
\sqrt{\alpha\theta}n^{(EB)}_{k} + n^{(B)}_{k}, & \mbox{Otherwise}.\\
\end{array},
\right.
\end{equation}
\end{small}
\hrule
\end{figure*}

From the model in \eqref{eq:eq_signal_model_with_attack}, it seems that Alice and Bob can circumvent the CA by employing encoding and decoding mechanisms that do not rely on the knowledge of channel state information (CSI), such as differential-encoding methods and blind detection techniques \cite{XGB, TsG}. However, these methods work under the assumption that some statistics of the channel remain constant for several blocks, and are also known at the receiver. In the case of CA, these techniques are not directly applicable as $w(t)$ is completely controlled by Eve. In the case of frequency-selective equivalent channel, the delayed components are contributed only by Eve as the main channel is frequency-flat due to the narrowband assumption. A straightforward way to handle frequency-selectivity is by using OFDM as the modulation scheme. However, the idea of OFDM modulation requires the channel realizations to be fixed for at least one OFDM symbol, and this assumption can also be violated by Eve. Therefore, OFDM is not applicable in this attack scenario.

\begin{figure}
\begin{center}
\includegraphics[scale=0.37]{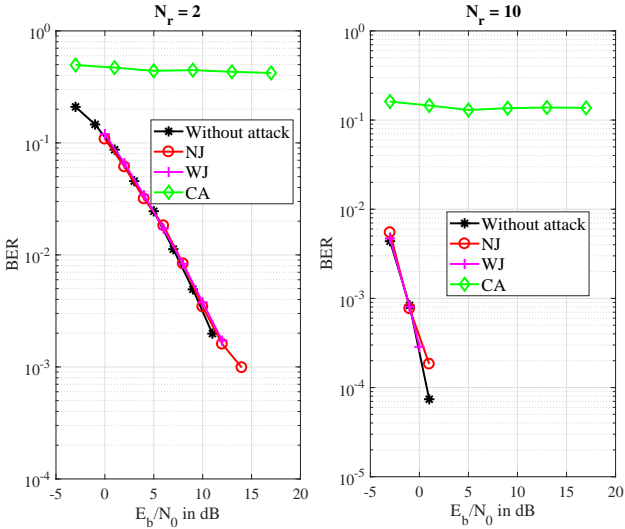}
\vspace{-0.7cm}
\caption{\label{fig:ca_attack_effect}Impact of CA, Narrowband Jamming and Wideband Jamming on an FH-based communication with $N = 1024$ carrier frequencies. Binary Phase Shift Keying (BPSK) signalling scheme is used at Alice aided by coherent maximum-likelihood detection at Bob}
\end{center}
\end{figure}

\subsection{Impact of Convolution Attack}
\label{subsec:nj_vs_ca}

To showcase the impact of CA, we consider a Binary Phase Shift Keying (BPSK) signalling scheme at Alice aided by coherent maximum-likelihood detection at Bob. We present the error-performance of this scheme under the following attacks: (i) \emph{Narrowband jamming (NJ)}: Eve executes narrowband jamming by injecting noise of energy $E_{eve} = \theta E_{Alice}$, with $\theta = 9$, on one of the $N = 1024$ bands with uniform distribution, (ii) \emph{Wideband Jamming (WJ)}: Eve executes wideband jamming by uniformly dividing its energy $E_{eve} = \theta E_{Alice}$, with $\theta = 9$, across the $N = 1024$ narrowbands, and (iii) \emph{Convolution attack (CA)}: Eve executes CA to result in a rapid-fading frequency-flat equivalent channel, with $\alpha = 1$ and $\theta = 9$. The equivalent channel on each carrier, denoted by $\sqrt{E_{Alice}}h^{(AB)}_{k} + \sqrt{E_{Alice}\alpha \theta}h^{(AEB)}_{k}$, changes rapidly to force error-floor behaviour on coherent maximum-likelihood detection. For the experiments, we use $\sigma^{2}_{Bob} = 1$ and $\sigma^{2}_{Eve} = 0.01$. In Fig. \ref{fig:ca_attack_effect}, we plot the bit-error-rate (BER) curves of the above schemes against $\frac{E_{b}}{N_{0}} = \frac{E_{Alice}}{2\sigma^{2}_{Bob}}$ when Bob is equipped with $N_{r} = 2$ and $N_{r} = 10$ receive antennas. The plots show that neither narrowband jamming nor wideband jamming is effective in degrading the error-performance at Bob, whereas CA can force severe BER degradation at an attack-ignorant Bob. Thus, even with large values of $N$, it is important for Alice and Bob to identify CA, and then mitigate it by employing an appropriate countermeasure.

\section{Convolution Attack on FH Based On-Off Keying}
\label{sec:on_off}

We study an FH system with non-coherent On-Off Keying (OOK) as the modulation scheme. In this strategy, Alice communicates bit-$1$ by transmitting a signal of energy $E_{Alice}$ (referred to as ON state), and bit-$0$ by switching-off the communication (referred to as OFF state). For exposition, let $b_{k}$ denote the bit transmitted at the $k$-th time instant. To communicate $b_{k}$, Alice encodes it as
\begin{equation}
\label{eq:on-off-keying}
x_{k} = \left\{ \begin{array}{cccccccccc}
1, & \mbox{ if } b_{k} = 1;\\
0, & \mbox{otherwise},\\
\end{array}
\right.
\end{equation}
before transmitting $x_{k}$ on the carrier-frequency $f_{k} \in \mathcal{F}$. From the nature of the attack, it is clear that Eve forwards only the noise component in the active narrowband when Alice switches-off her transmitter. However, when Alice transmits bit-$1$, Eve forwards significant signal power in the active narrowband. With this signal design, Bob can distinguish bit-$1$ and bit-$0$ by measuring the received energy on each symbol without the knowledge of the channel. In the rest of this paper, we assume that Eve uses $w(t)$ which results in a frequency-flat equivalent channel at Bob. For the frequency-selective case, Alice and Bob may handle it by locking onto a carrier-frequency $f_{c}$ only for one symbol, i.e., $m = 1$, so that Bob may continue to listen to the preceding set of carrier-frequencies for the delayed components. Our inferences on the attack-strategies and countermeasures are only based on the frequency-flat equivalent channel model. Since the users can handle frequency-selectivity by hopping across the carrier-frequencies for one symbol, we do not expect significant deviations in the inferences with the frequency-selective case.

Applying OOK on the frequency-flat model, the received symbol at Bob is given in \eqref{eq:on-off-keying_rx_signal} (top of this page), where $\sqrt{\alpha \theta}$ is the gain applied by Eve on its received signal. Based on the nature of operations at Eve, we model the complex channel $h^{(AEB)}_{k}$ as $h^{(AEB)}_{k} \triangleq h^{(AE)}_{k}w_{k}h^{(EB)}_{k}$, where $h^{(AE)}_{k}$ is the channel from Alice to Eve, distributed as $\mathcal{CN}(0, 1)$, $h^{(EB)}_{k}$ is the channel from Eve to Bob, distributed as $\mathcal{CN}(0, 1)$, and $w_{k}$ is a complex random variable of mean zero and unit variance obtained from the waveform $w(t)$. The forwarded additive noise from Eve is $n^{(EB)}_{k} \triangleq h^{(EB)}_{k}w_{k}n^{(E)}_{k}$, where $n^{(B)}_{k}$ is distributed as $\mathcal{CN}(0, \sigma^{2}_{Eve})$. Henceforth, we denote $\alpha \theta E_{Alice}$ as $E_{Eve, C}$, which is the additional signal energy contributed by Eve through CA.

At the receiver side, Bob decodes to an estimate of $b_{k}$, denoted by $\hat{b}_{k}$, based on the following rule:
\begin{equation}
\label{eq:on-off-keying_decoding}
\hat{b}_{k} = \left\{ \begin{array}{cccccccccc}
1, & \mbox{ if } |y_{k}|^2 > E_{th};\\
0,  & \mbox{Otherwise}.\\
\end{array},
\right.
\end{equation}
where $E_{th}$ is an appropriately designed threshold chosen based on the noise component in \eqref{eq:on-off-keying_rx_signal}. One of the challenges in designing OOK against the CA is the derivation of the threshold $E_{th}$, as fading characteristics of $h^{(AB)}_{k}$ and $h^{(AEB)}_{k}$ have to be considered. We address the choice of $E_{th}$ in Section \ref{sec:threshold_choice}. 

When decoding OOK, Bob faces two types of error events: (i) $\hat{b}_{k} = 1$ when $b_{k} = 0$, and (ii) $\hat{b}_{k} = 0$ when $b_{k} = 1$. While the former event may occur when the threshold $E_{th}$ is lower than the noise components jointly contributed by Eve and Bob, the latter event captures the case when Eve attempts to force the effective channel $\sqrt{E_{Alice}}h^{(AB)}_{k} + \sqrt{\alpha\theta E_{Alice}}h^{(AEB)}_{k}$ to deep fade, i.e., $|y_{k}|^2 \leq E_{th}$. We represent the associated probability as $P^{(attack)}_{1 \rightarrow 0}$. In the following section, we propose a mitigation strategy by Bob to reduce $P^{(attack)}_{1 \rightarrow 0}$.

\subsection{Mitigation Strategy: Large Number of Receive Antennas}
\label{subsec:cross_1_0}


In the case of CA, since $w(t)$ is completely controlled by Eve, the distribution of the equivalent channel can be changed to affect $P^{(attack)}_{1 \rightarrow 0}$ provided $E_{Eve, C} >> E_{Alice}$. However, on the defense-side, since the two users hop across a wide range of narrowbands, Eve cannot learn the narrowband, and therefore, she cannot drive the equivalent channel $\sqrt{E_{Alice}}h^{(AB)}_{k} + \sqrt{E_{Eve, C}}h^{(AEB)}_{k}$ to deep fade with probability one. As a defense mechanism to counter Eve's strategy, Bob should collect energy from as many independent paths as possible. One such bandwidth-efficient way is to employ multiple receive antennas at Bob. This way, the probability that Eve can drive all the independent channels simultaneously to deep fade can be reduced. If we use $N_{r}$ to denote the number of receive antennas at Bob, without additive-noise at Eve and Bob, the total signal energy collected across $N_{r}$ antennas is given by
\begin{equation}
\label{eq:rx_energy_with_attack}
R^{(attack)}_{N_{r}, k} \triangleq \sum_{j = 1}^{N_{r}} \left|\sqrt{E_{Alice}}h^{(AB)}_{k, j} + \sqrt{E_{Eve, C}}h^{(AE)}_{k}w_{k}h^{(EB)}_{k, j}\right|^2,
\end{equation}
where $h^{(AB)}_{k, j}$ and $h^{(EB)}_{k, j}$ denote the equivalent channels seen by the $j$-th antenna of Bob on the $k$-th symbol. In the event of no attack, we have $w_{k} = 0$, and $R^{(no-attack)}_{N_{r}, k}$, given by
\begin{equation}
\label{eq:rx_energy_without_attack}
R^{(no-attack)}_{N_{r}, k} \triangleq \sum_{j = 1}^{N_{r}} \left|\sqrt{E_{Alice}}h^{(AB)}_{k, j}\right|^2,
\end{equation}
is Chi-square distributed with degrees of freedom $2N_{r}$. However, with attack, the error-performance depends on the distribution of $R^{(attack)}_{N_{r}, k}$ given in \eqref{eq:rx_energy_with_attack}, which in turn depends on the distribution of $w_{k}$. When $N_{r}$ is large, the following proposition shows that Eve's additional energy can be used to Bob's advantage to accumulate more energy. Although this result seems to suggest that CA is aiding Bob to improve the error-performance, it is important to note that this relative improvement is with respect to non-coherent OOK, which is already sub-optimal compared to coherent ML detection techniques.

\begin{small}
\begin{figure*}
\begin{small}
\begin{eqnarray}
\label{eq:expand_energy}
R^{(attack)}_{N_{r}, k} = \sum_{j = 1}^{N_{r}} \left( \left|\sqrt{E_{Alice}}h^{(AB)}_{k, j}\right|^2 + \left|\sqrt{E_{Eve, C}}h^{(AE)}_{k}w_{k}h^{(EB)}_{k, j}\right|^2\right) + \sum_{j = 1}^{N_{r}} \left( \sqrt{E_{Alice}E_{Eve, C}} 2 \mathcal{R}\left(h^{*(AB)}_{k, j}h^{(AE)}_{k}w_{k}h^{(EB)}_{k, j}\right) \right)
\end{eqnarray}
\end{small}
\begin{footnotesize}
\begin{eqnarray}
\label{eq:diff_energy}
\mbox{Prob}\left(\frac{R_{\Delta}}{N_{r}} > -\epsilon\right) = 
\mbox{Prob}\left(\left(\frac{1}{N_{r}} \left(\sum_{j = 1}^{N_{r}} \left|\sqrt{E_{Eve, C}} h^{(AE)}_{k}w_{k}h^{(EB)}_{k, j}\right|^2 +  
 \sum_{j = 1}^{N_{r}} \left( \sqrt{E_{Eve, C}E_{Alice}} 2 \mathcal{R}\left(h^{*(AB)}_{k, j}h^{(AE)}_{k}w_{k}h^{(EB)}_{k, j}\right) \right)\right) \right) > - \epsilon \right)
\end{eqnarray}
\end{footnotesize}
\begin{small}
\begin{eqnarray}
\label{eq:diff_energy_lower_bound}
\mbox{Prob}\left(\frac{R_{\Delta}}{N_{r}} > -\epsilon\right) & \geq & \mbox{Prob}\left(\left|\frac{1}{N_{r}} \sum_{j = 1}^{N_{r}} \left( \sqrt{E_{Alice}E_{Eve, C}} 2 \mathcal{R}\left(h^{*(AB)}_{k, j}h^{(AE)}_{k}w_{k}h^{(EB)}_{k, j}\right) \right)\right| < \epsilon \right)\\
\label{eq:diff_energy_lower_bound1}
& = & \mbox{Prob}\left(\left|\left(\frac{1}{N_{r}} \sum_{j = 1}^{N_{r}} \left( \sqrt{E_{Alice}E_{Eve, C}} 2 \mathcal{R}\left(h^{*(AB)}_{k, j}h^{(AE)}_{k}w_{k}h^{(EB)}_{k, j}\right) \right)\right) - 0 \right| < \epsilon \right)\\
\label{eq:diff_energy_lower_bound2}
& > & 1 - \epsilon.
\end{eqnarray}
\end{small}
\hrule
\end{figure*}
\end{small}

\begin{Proposition}
\label{prop:large_antenna}
Let $R_{\Delta} \triangleq R^{(attack)}_{N_{r}, k} - R^{(no-attack)}_{N_{r}, k}$, where $R^{(attack)}_{N_{r}, k}$ and $R^{(attack)}_{N_{r}, k}$ are as given in \eqref{eq:rx_energy_with_attack} and \eqref{eq:rx_energy_without_attack}, respectively. For a small $\epsilon > 0$, there exists $\bar{N}_{r}$ such that for all $N_{r} \geq \bar{N}_{r}$, we have
\begin{equation}
\label{eq:theorem_result_diff_energy}
\mbox{Prob}\left(\frac{R_{\Delta}}{N_{r}} > -\epsilon\right) > 1 - \epsilon.
\end{equation}
\end{Proposition}
\begin{IEEEproof}
We start by expanding $R^{(attack)}_{N_{r}, k}$ as in \eqref{eq:expand_energy}, where $\sum_{j = 1}^{N_{r}} |\sqrt{E_{Alice}}h^{(AB)}_{k, j}|^2$ is the energy accumulated at Bob without the attack. 
As a result, $\mbox{Prob}\left(\frac{R_{\Delta}}{N_{r}} > -\epsilon\right)$ can be written as in \eqref{eq:diff_energy}. Since $\sum_{j = 1}^{N_{r}} |\sqrt{E_{Eve, C}}h^{(AE)}_{k}w_{k}h^{(EB)}_{k, j}|^2$ is strictly non-negative, the probability in \eqref{eq:diff_energy} is lower-bounded by \eqref{eq:diff_energy_lower_bound}. This is because we are only considering the events when $$\frac{1}{N_{r}} \sum_{j = 1}^{N_{r}} \left( \sqrt{E_{Alice}E_{Eve, C}} 2 \mathcal{R}\left(h^{*(AB)}_{k, j}h^{(AE)}_{k}w_{k}h^{(EB)}_{k, j}\right) \right)$$
is bounded in the interval $(-\epsilon, \epsilon)$. Furthermore, the random variables $\left\lbrace \mathcal{R}\left(h^{*(AB)}_{k, j}h^{(AE)}_{k}w_{k}h^{(EB)}_{k, j}\right) ~|~ 1 \leq j \leq N_{r}\right\rbrace$ are i.i.d. with mean zero since $h^{(AE)}_{k}$ and $w_{k}$ are constants. As a result, we rewrite \eqref{eq:diff_energy_lower_bound} as \eqref{eq:diff_energy_lower_bound1}. Finally, applying weak law of large numbers \cite[Chapter 3]{info_theory} on \eqref{eq:diff_energy_lower_bound1}, we get \eqref{eq:diff_energy_lower_bound2} for sufficiently large $N_{r} \geq \bar{N}_{r}$. This completes the proof.
\end{IEEEproof}

With massive MIMO in contention for next-generation networks (e.g. 5G), base-stations equipped with hundreds of antennas are likely to be deployed in practice \cite{massive_mimo}. This implies that Proposition \ref{prop:large_antenna} is useful when base-station plays the role of Bob and a UE (user-equipment) plays the role of Alice.

\begin{figure}
\begin{center}
\includegraphics[scale=0.4]{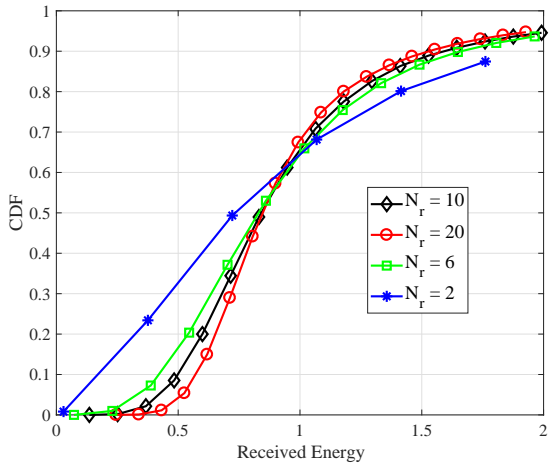}
\vspace{-0.3cm}
\caption{\label{fig:cdf_Gaussian_SIMO}CDFs of the average received energy across $N_{r}$ antennas at Bob. The parameter $\eta$, as given in \eqref{eq:eta_definition}, denotes the percentage of energy contributed by Eve at Bob. The plots show that multiple receive antennas at Bob helps to reduce the attack effect.}
\end{center}
\end{figure}

While the above proposition shows the advantage of employing large number of receive antennas to combat the CA, in the rest of this section, we present numerical results to understand the cumulative distribution function (CDF) of $R^{(attack)}_{N_{r}, k}$ when $N_{r}$ is not large. To generate the numerical results, we assume that the channels $h^{(AB)}_{k, j}, h^{(AE)}_{k}$ and $h^{(EB)}_{k, j}$ are i.i.d., and are distributed as $\mathcal{CN}(0, 1)$.  We also assume that $w_{k}$ is distributed as $\mathcal{CN}(0, 1)$. When the transmitted bit is $1$, let $E_{total} = E_{Alice} + E_{Eve, C}$ be the average received energy at every antenna of Bob, out of which, $E_{Eve, C}$ be the signal energy contributed by Eve. We define 
\begin{equation}
\label{eq:eta_definition}
\eta \triangleq \frac{E_{Eve, C}}{E_{total}} \times 100,
\end{equation}
as the percentage of average energy contributed by Eve when the transmitted bit is $1$. In Fig. \ref{fig:cdf_Gaussian_SIMO}, we plot the CDFs of the random variable $\frac{R^{(attack)}_{N_{r}, k}}{N_{r}}$ when $w_{k}$ is Gaussian distributed. For computing the CDFs, we use $E_{total} = 1$. The plots in Fig. \ref{fig:cdf_Gaussian_SIMO} show that as $N_{r}$ increases, the CDFs shift towards right, thereby driving the cross-over probability to lower values.

\begin{figure*}
\begin{equation}
\label{eq:energy_on_state}
E_{bit-1} = \sum_{j = 1}^{N_{r}} \left|\sqrt{E_{Alice}}h^{(AB)}_{k, j} + \sqrt{E_{Eve, C}}h^{(AE)}_{k}w_{k}h^{(EB)}_{k, j} + \sqrt{\alpha \theta}n^{(EB)}_{k, j} + n^{(B)}_{k, j} \right|^2
\end{equation}
\begin{equation}
\label{eq:approximate_energy_on_state}
\tilde{E}_{bit-1} = \sum_{j = 1}^{N_{r}} \left|\sqrt{E_{Alice}}h^{(AB)}_{k, j} + \sqrt{E_{Eve, C}}\tilde{h}^{(AEB)}_{k, j} + \sqrt{\alpha \theta}\tilde{n}^{(EB)}_{k, j} + n^{(B)}_{k, j}\right|^2
\end{equation}
\begin{equation}
\label{eq:approximate_energy_off_state}
\tilde{E}_{bit-0} = \sum_{j = 1}^{N_{r}} \left|\sqrt{\alpha \theta}\tilde{n}^{(EB)}_{k, j} + n^{(B)}_{k, j}\right|^2
\end{equation}
\hrule
\end{figure*} 

\subsection{Effect of Multiple antennas at Eve}
\label{subsec:mult_Eve}

\begin{figure}
\begin{center}
\includegraphics[scale=0.4]{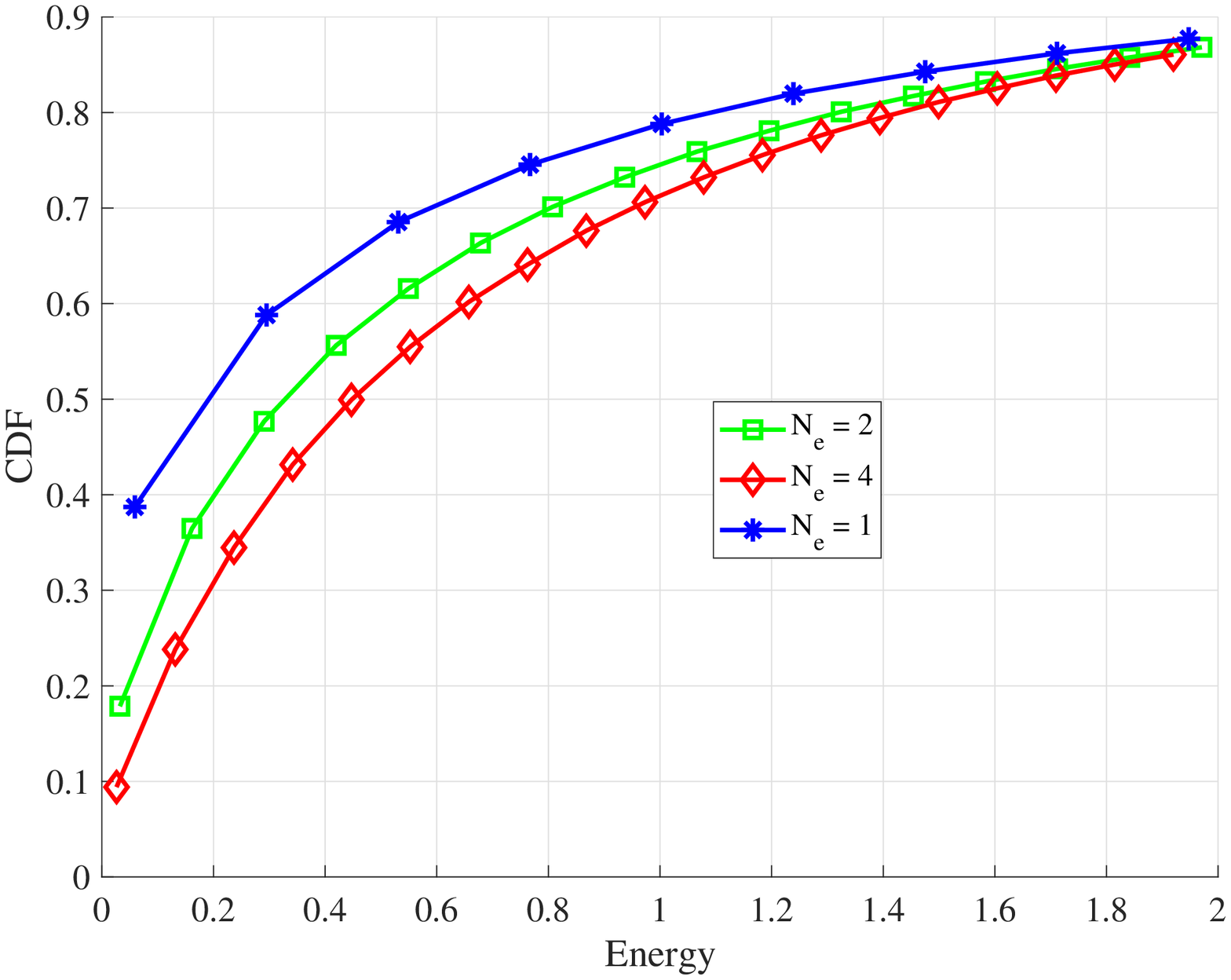}
\caption{\label{fig:CDF} CDFs of $|\cdot|^{2}$ of random variables in \eqref{sum_prod}, where $|\cdot|$ denotes the absolute value of a complex number. The plots indicate that using multiple antennas at Eve changes the energy distribution at Bob.}
\end{center}
\end{figure}

We acknowledge that Bob's trick to garner energy for detection comes from using multiple antennas. To keep the comparison fair, we study the effect of CA when Eve is also equipped with multiple antennas. Considering $N_{r} = 1$, the total energy at Bob without additive noise at Alice and Bob is given by
\begin{equation}
\label{eq:energy_multiple_eve}
\left|\sqrt{E_{Alice}} h^{(AB)}_{k} + \sum_{l = 1}^{N_{e}} \left(\sqrt{\frac{\alpha \theta E_{Alice}}{N_{e}}}\right)h^{(EB)}_{k, l} h^{(AE)}_{k, l} w_{k, l}\right|^2,
\end{equation}
where $N_{e}$ denotes the number of antennas at Eve, $w_{k, l}$, which is distributed as $\mathcal{CN}(0, 1)$, is the scalar used at the $l$-th antenna of Eve, $h^{(EB)}_{k, l}$ is the channel from the $l$-th antenna at Eve to Bob, and $h^{(AE)}_{k, l}$ is the channel from Alice to the $l$-th antenna at Eve. In the case of single antenna at Eve, the energy at Bob is
\begin{equation}
\label{eq:energy_single_eve}
\left|\sqrt{E_{Alice}} h^{(AB)}_{k} + \sqrt{\alpha \theta E_{Alice}}h^{(EB)}_{k, 1} h^{(AE)}_{k, 1} w_{k, 1}\right|^2
\end{equation}
where the main difference between \eqref{eq:energy_multiple_eve} and \eqref{eq:energy_single_eve} is the distribution of the random variables 
\begin{equation}
\label{sum_prod}
\sum_{l = 1}^{N_{e}} \frac{1}{\sqrt{N_{e}}}h^{(EB)}_{k, l} w_{k, l} h^{(AE)}_{k, l}
\end{equation}
with $N_{e} > 1$ and with $N_{e} = 1$.
With $N_{e} > 1$, since \eqref{sum_prod} is the sum of product of three independent Gaussian random variables, we have observed that the CDF of $\left|\sum_{l = 1}^{N_{e}} \frac{1}{\sqrt{N_{e}}}h^{(EB)}_{k, l} w_{k, l} h^{(AE)}_{k, l}\right|^2$ grows much slower than that of $\left|h^{(EB)}_{k, 1} w_{k, 1} h^{(AE)}_{k, 1}\right|^2$, as shown in Fig. \ref{fig:CDF}. As a result, for a given $E_{th}$, the probability of decoding bit-$1$ as bit-$0$ decreases when multiple antennas are used at Eve.

\begin{figure}
\begin{center}
\includegraphics[scale=0.37]{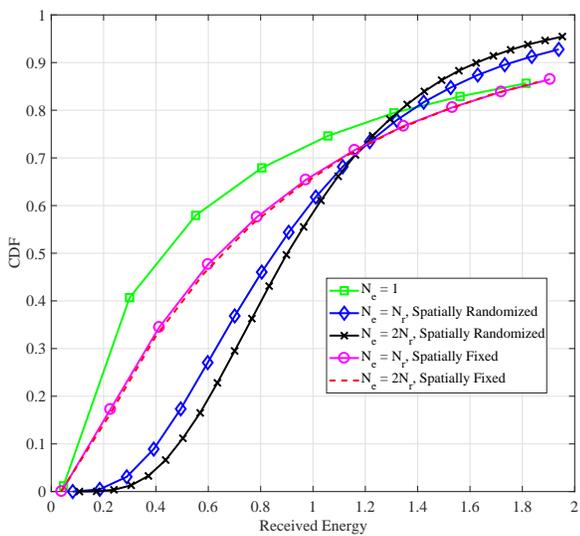}
\vspace{-0.5cm}
\caption{\label{fig:cdf_overall}Comparison of CDFs of the average received energy across $N_{r}$ antennas at Bob for various strategies employed at Eve. For the results, we fix $N_{r} = 10$ and $\eta = 90\%$. The number of antennas at Eve is denoted by $N_{e}$. The plots highlight that it is better for Eve to equip only one antenna in order increase the attack effect.}
\end{center}
\end{figure}

Furthermore, we compute the CDFs of the received energy across the $N_{r}$ antennas at Bob when $N_{r} \geq 1$ and Eve uses the following strategies: (i) single-antenna, (ii) multiple-antenna with spatially randomized waveforms  - $\{w_{k, l} ~|~ 1 \leq l \leq N_{e}\}$ are independent, and (iii) multiple-antenna with spatially fixed waveforms - $\{w_{k, l} = w_{k} ~|~ 1 \leq l \leq N_{e}\}$. To give advantage to Eve, we have also considered the case when $N_{e} = 2N_{r}$. The CDFs, which are presented in Fig. \ref{fig:cdf_overall}, highlight that employing multiple antennas at Eve does not aggravate the attack effect as multiple antennas assists Bob in receiving more energy than the single-antenna case. Due to lack of closed-form expressions on the CDFs of energy collected at Bob, we do not have concrete theoretical insights on this argument. Nevertheless, based on the simulation results, we advocate the use of single antenna at Eve and multiple antennas at Bob. In Section \ref{sec:sims}, we also present the BER performance of OOK with and without multiple antennas at Eve to reinforce this observation.

\begin{figure}
\begin{center}
\includegraphics[scale=0.32]{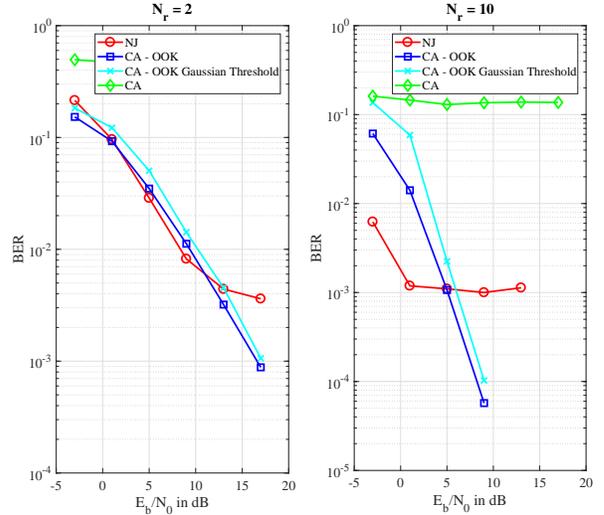}
\vspace{-0.3cm}
\caption{\label{fig:ber_N_128_Nr_2}Error-performance of OOK against convolution attack (CA) on an FH system with $N = 128$. Since the attack is persistent, Bob measures the energy distributions during the attack to design the threshold $E^{*}_{th}$. The choice of $\tilde{E}^{*}_{th}$, which is based on Gaussian approximation marginally degrades the performance compared to that when using the optimal value $E^{*}_{th}$.}
\end{center}
\end{figure} 

\begin{figure}
\begin{center}
\includegraphics[scale=0.34]{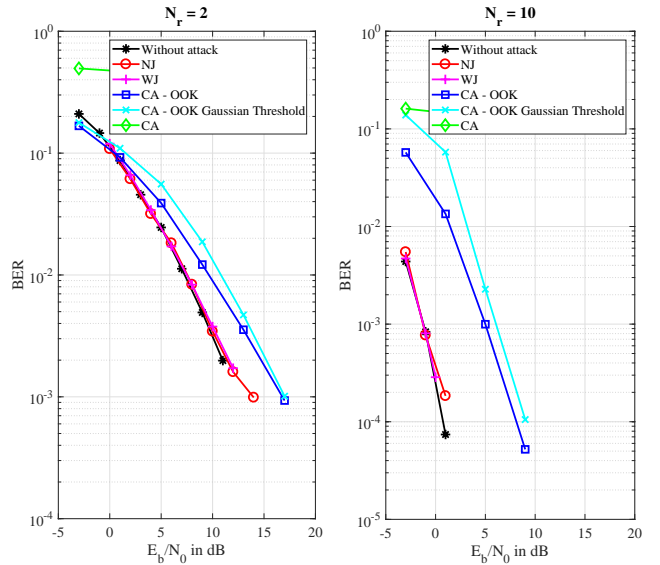}
\vspace{-0.3cm}
\caption{\label{fig:ber_nr_2}Error-performance of OOK against convolution attack (CA) on an FH system with $N = 1024$. Since the attack is persistent, Bob measures the energy distributions during the attack to design the thresholds $E^{*}_{th}$ and $\tilde{E}^{*}_{th}$.}
\end{center}
\end{figure}

\subsection{Design of Threshold $E_{th}$}
\label{sec:threshold_choice}

Having studied the energy distributions during the ON state of OOK, we now address the computation of $E_{th}$ in \eqref{eq:on-off-keying_decoding} to optimize the error-performance at Bob. With CA, the signal energy collected across $N_{r}$ antennas during the ON state is given by \eqref{eq:energy_on_state} (see the top of the next page),
where $h^{(AB)}_{k, j}$ and $h^{(EB)}_{k, j}$ denote the equivalent channels seen by the $j$-th antenna of Bob on the $k$-th symbol. Similarly, energy collected during the OFF state is
\begin{equation}
E_{bit-0} = \sum_{j = 1}^{N_{r}} \left|\sqrt{\alpha \theta}n^{(EB)}_{k, j} + n^{(B)}_{k, j}\right|^2.
\end{equation}
To determine the optimal threshold we need to solve
\begin{equation}
\label{eq:optimal_optimization}
E^{*}_{th} = \arg min_{E_{th}} \mbox{Prob}(E_{bit-1} \leq E_{th}) + \mbox{Prob}(E_{bit-0} > E_{th}),
\end{equation}
which in turn requires Bob to measure the Probability Density Functions (PDFs) on $E_{bit-1}$ and $E_{bit-0}$. Towards that direction, we assume that Bob can learn the distributions empirically using pilots, which are periodically transmitted by Alice. Note that the persistent nature of the CRFH attack helps in measuring the energy distributions with attack, otherwise, Bob is forced to employ threshold values based on the energy distribution of $\sqrt{E_{Alice}}h^{(AB)}_{k}$, which in turn degrades the error-performance under convolution attack. From $E_{bit-1}$ and $E_{bit-0}$, we observe that $\{h^{(AE)}_{k}w_{k}h^{(EB)}_{k, j}~|~ 1\leq j\leq N_{r}\}$ are uncorrelated but not necessarily independent. Similarly, the random variables $\{n^{(EB)}_{k, j}~|~ 1\leq j\leq N_{r}\}$ are also uncorrelated but not independent. Due to challenges in obtaining the closed-form expressions on the PDFs of $E_{bit-1}$ and $E_{bit-0}$, we approximate $\{h^{(AE)}_{k}w_{k}h^{(EB)}_{k, j}~|~ 1\leq j\leq N_{r}\}$ to be statistically independent and Gaussian distributed as $\mathcal{CN}(0, 1)$, and then arrive at a sub-optimal solution. Similarly, $\{n^{(EB)}_{k, j}~|~ 1\leq j\leq N_{r}\}$ is also assumed i.i.d., where each $n^{(EB)}_{k, j}$ is distributed as $\mathcal{CN}({0, \sigma^{2}_{Eve}})$. Using such approximations, the corresponding versions of received energy are given by \eqref{eq:approximate_energy_on_state} and \eqref{eq:approximate_energy_off_state},
where $\tilde{h}^{(AEB)}_{k, j}$ and $\tilde{n}^{(EB)}_{k, j}$ are Gaussian distributed. We immediately note that $\tilde{E}_{bit-1}$ can be written as 
\begin{equation*}
\tilde{E}_{bit-1} = \frac{1}{2}(E_{Alice} + E_{Alice}\alpha \theta + \alpha \theta \sigma_{Eve}^{2} + \sigma_{Bob}^{2})\chi_{1}
\end{equation*}
where $\chi_{1}$ is Chi-square distributed with degrees of freedom $2N_{r}$. Similarly, $\tilde{E}_{bit-0}$ can be written as 
\begin{equation*}
\tilde{E}_{bit-0} = \frac{1}{2}(\alpha \theta \sigma_{Eve}^{2} + \sigma_{Bob}^{2})\chi_{2}
\end{equation*}
where $\chi_{2}$ is also a Chi-square distributed random variable with degrees of freedom $2N_{r}$. With this, the approximate solution, henceforth denoted as $\tilde{E}^{*}_{th}$, is computed as in \eqref{eq:approximate_optimization},
\begin{figure*}
\begin{eqnarray}
\label{eq:approximate_optimization}
\tilde{E}^{*}_{th} & = & \arg min_{E_{th}}  \mbox{Prob}\left(\chi_{1} \leq \frac{2E_{th}}{E_{Alice} + E_{Alice}\alpha \theta + \alpha \theta \sigma_{Eve}^{2} + \sigma_{Bob}^{2}}\right) + \mbox{Prob}\left(\chi_{2} > \frac{2E_{th}}{\alpha \theta \sigma_{Eve}^{2} + \sigma_{Bob}^{2}}\right)\nonumber\\
& = & \arg min_{E_{th}}  \gamma \left(N_{r}, \frac{E_{th}}{E_{Alice} + E_{Alice}\alpha \theta \sigma_{Eve}^{2} + \alpha \theta + \sigma_{Bob}^{2}}\right) - \gamma \left(N_{r}, \frac{E_{th}}{\alpha \theta \sigma_{Eve}^{2} + \sigma_{Bob}^{2}}\right)
\end{eqnarray}
\hrule
\end{figure*}
wherein $\gamma(\cdot, \cdot)$ is the lower incomplete gamma function. Unlike the optimal solution in \eqref{eq:optimal_optimization}, the solution in \eqref{eq:approximate_optimization} can be obtained using numerical methods on incomplete gamma function.

\subsection{Error-Performance of OOK Against Convolution Attack}
\label{sec:sims}

In this section, we present simulation results on the error-performance of OOK against the CA. To carry out the experiments, we assume that the channels $\{h^{(AB)}_{k, j}(f_{c}) ~|~ f_{c} \in \mathcal{F}\}$ across the $N$ narrowbands are statistically independent and distributed as $\mathcal{CN}(0, 1)$. Similarly, the sets of channels $\{h^{(AE)}_{k}(f_{c}) ~|~ f_{c} \in \mathcal{F}\}$ and $\{h^{(EB)}_{k, j}(f_{c}) ~|~ f_{c} \in \mathcal{F}\}$ are also i.i.d. across the $N$ narrowbands, and are distributed as $\mathcal{CN}(0, 1)$. 


To showcase the effect of CA, we present the error-performance of the non-coherent OOK scheme along with the schemes discussed in Section \ref{subsec:nj_vs_ca}, namely, (i) Narrowband jamming (NJ), and (ii) the CA, on binary phase shift keying (BPSK) with coherent maximum-likelihood detection at Bob.
In Fig. \ref{fig:ber_N_128_Nr_2}, we plot the BER curves of the above schemes against $\frac{E_{b}}{N_{0}} = \frac{E_{Alice}}{2\sigma^{2}_{Bob}}$ for $N_{r} = 2$ and $N_{r} = 10$. For CA on OOK, we use two different threshold values for energy detection, namely: $E^{*}_{th}$ in \eqref{eq:optimal_optimization}, and $\tilde{E}^{*}_{th}$ in \eqref{eq:approximate_optimization}, which are computed based on the attack parameters. The plots show that the Gaussian approximation to compute $\tilde{E}^{*}_{th}$ does not result in significant loss in the error-performance. Moreover, the error-performance of OOK is better than that of coherent modulation method under the CA. Similar to the results in Fig. \ref{fig:ber_N_128_Nr_2}, we also present the BER curves of OOK with $N = 1024$ in Fig. \ref{fig:ber_nr_2}. The plots highlight that it is important for Alice and Bob to identify the CA, and then mitigate it by employing OOK based strategy. 

\begin{figure}[h]
\begin{center}
\includegraphics[scale=0.34]{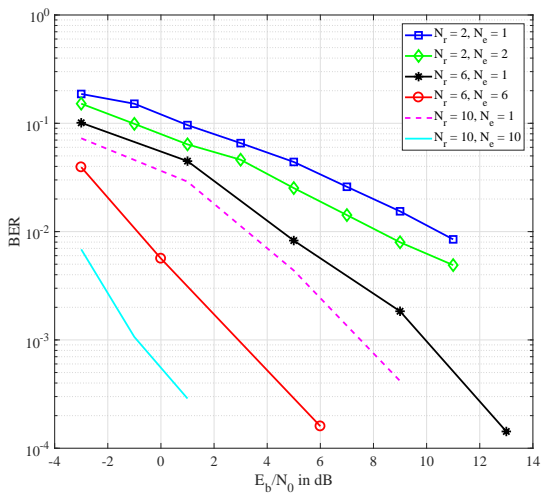}
\vspace{-0.4cm}
\caption{\label{fig:ber_mimo_effect}Error-performance of OOK against CA when Eve is equipped with multiple antennas. The plots show that the best attack strategy for Eve is to mount just one antenna.}
\end{center}
\end{figure}

Finally, in Fig. \ref{fig:ber_mimo_effect}, we present the error-performance of OOK when Eve is equipped with multiple antennas, and when $\theta = 9, \alpha = 100 \%$ and $N = 1024$. Similar to the observations in Section \ref{subsec:mult_Eve}, Fig. \ref{fig:ber_mimo_effect} confirms that multiple antennas at Eve does not aggravate the attack effect. For the simulations, the threshold values for energy detection are computed based on the energy distribution during the ON and the OFF states similar to the one in \eqref{eq:optimal_optimization}.

\begin{remark}
The error-performance of OOK, as presented in Fig. \ref{fig:ber_N_128_Nr_2} and Fig. \ref{fig:ber_nr_2}, captures the best-case results from the perspective of Alice and Bob. This is attributed to the assumption that Eve executes the convolution attack persistently on both the pilot symbols and the data symbols with the same parameters $\alpha$ and $\theta$. However, when Eve selectively attacks only the data symbols, then the corresponding estimate of the threshold will be suboptimal, which in turn will result in degraded performance.
\end{remark}

\subsection{Limitations of OOK against Wideband Jamming}

\begin{figure}
\begin{center}
\includegraphics[scale=0.33]{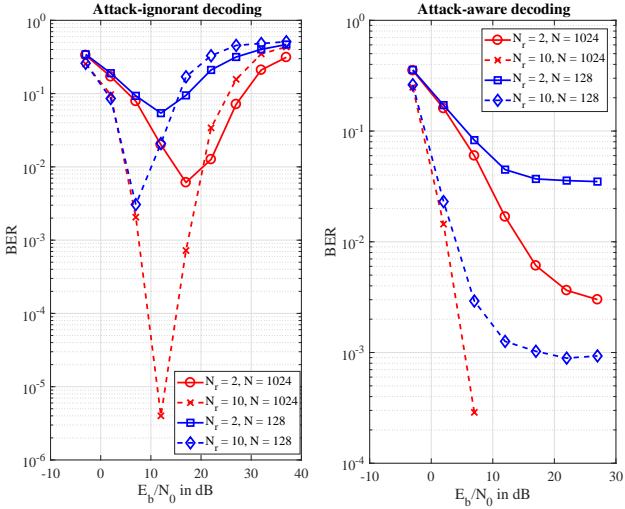}
\vspace{-0.4cm}
\caption{\label{fig:ber_ook_wideband_jamming} Error-performance of OOK against wideband jamming on an FH system with $N = 128$ and $N = 1024$. We use $\theta = 9$ to generate the results.}
\end{center}
\end{figure}

In this section, we explore the idea of changing Eve's strategy to wideband jamming (WJ) once Alice and Bob switch to OOK in response to CA. In WJ, Eve uniformly divides her energy $E_{Eve} = \theta E_{Alice}$ across the $N$ narrowbands. The rationale behind this switch is to exploit lower threshold values used for energy detection, thereby forcing Bob to decode bit-$0$ as bit-$1$. We first capture the consequence of an attack-ignorant detection in Fig. \ref{fig:ber_ook_wideband_jamming} (left-side), which shows the BER performance of OOK when $E_{th}$ is optimized based on $E_{Alice}$ and $\sigma^{2}_{Bob}$. Since $E_{th}$ is chosen based on $E_{Alice}$, and $E_{Alice}$ is much lower than $E_{Eve}$, BER increases with large values of jamming energy; this is mainly contributed by the error event of decoding bit-$0$ as bit-$1$. In the attack-aware case, Bob measures the jamming energy using the pilots, and then takes it into account when designing $E_{th}$ (this is possible due to the persistent nature of the attack). The error-performance of such a strategy is also presented in Fig. \ref{fig:ber_ook_wideband_jamming} (right-side), which shows that unlike the case of attack-ignorant detection, the BER experiences error-floor behaviour when $E_{b}/N_{o}$ is large; this is because the threshold value linearly increases with $E_{Alice}$, thereby saturating the probability of decoding bit-$0$ as bit-$1$, and vice versa. In summary, although OOK mitigates CA, a combination of CA followed by WJ can result in degraded error-performance at Bob when $E_{b}/N_{o}$ is large. 

\section{Convolution Attack on FH based Frequency Shift Keying}
\label{sec:CA_on_FSK}

In this section, we study the impact of CA on Binary-FSK (BFSK) based FH scheme as an alternate countermeasure. We have chosen BFSK as the modulation scheme as most military and commercial frequency hopping systems use frequency shift keying. Unlike the generic attack in Section \ref{sec:CRFH_FH}, the objective of CRFH in this case is to create confusion at Bob when decoding the BFSK modulated symbols. In this attack, Eve uses an appropriate baseband signal $w(t)$ to forward a frequency-shifted version of the received passband signal so that the tones carrying bit-$0$ and bit-$1$ have comparable energy levels at Bob.

\subsection{Signal Model for BFSK without Attack}
\label{sec:CA_on_FSK_WOA}

At Alice, bit-$1$ is transmitted by using the carrier-frequency $f_{c} + \beta$, and bit-$0$ is transmitted by using the carrier-frequency $f_{c} - \beta$, for some $f_{c} \in \mathcal{F}$ (given in Section \ref{sec:CRFH_FH}). We assume that $0 < \beta < \frac{\Delta}{2}$, where $\Delta$ is the spacing between adjacent carrier-frequencies. We use $b_{k} \in \{0, 1\}$ to denote the bit transmitted at the $k$-th symbol-period, and  $\bar{b}_{k}$ to denote the complement of $b_{k}$. To communicate $b_{k}$, Alice transmits the tone $f_{k}$, given by
\begin{equation}
\label{eq:on-off-keying}
f_{k} = \left\{ \begin{array}{cccccccccc}
f_{c} + \beta, & \mbox{ if } b_{k} = 1,\\
f_{c} - \beta, & \mbox{otherwise},\\
\end{array}
\right.
\end{equation}
where $f_{c}$ is chosen based on the shared secret-key between Alice and Bob. Overall, the total set of tones used by Alice and Bob is $\mathcal{T} = \{f_{i} + \beta, f_{i} - \beta, ~|~ i = 1, 2, \ldots, N \}$. In this model, we assume $N_{e} = 1$ and $N_{r} = 1$.

Without any attack, the received complex-baseband symbols at Bob are of the form
\begin{eqnarray}
\label{eq:FSK_rx_signal_without_attack}
y_{k, main} & = & \sqrt{E_{Alice}}h^{(AB)}_{k} + n^{(B)}_{k, main}, \\
y_{k, side} & = & n^{(B)}_{k, side},
\end{eqnarray}
where $y_{k, main}$ and $y_{k, side}$ are the symbols received on the tones $f_{c} + \beta$ and $f_{c} - \beta$ depending on $b_{k}$. When bit-$1$ is transmitted, $y_{k, main}$ and $y_{k, side}$ correspond to the symbols on the tones $f_{c} + \beta$ and $f_{c} - \beta$, respectively. Similarly, when bit-$0$ is transmitted, $y_{k, main}$ and $y_{k, side}$ correspond to the symbols on the tones $f_{c} - \beta$ and $f_{c} + \beta$, respectively. Here, the additive white Gaussian noise (AWGN) components $n^{(B)}_{k, main}$ and $n^{(B)}_{k, side}$ are i.i.d. as $\mathcal{CN}(0, \sigma^2_{Bob})$. Stitching the above together, the received symbols on the tones $f_{c} + \beta$ and $f_{c} - \beta$ are given by 
\begin{equation}
\label{eq:main_carrier_mapping_1}
y_{k}(f_{c} + \beta) = \left\{ \begin{array}{cccccccccc}
y_{k, main}, & \mbox{ if } b_{k} = 1;\\
y_{k, side}, & \mbox{otherwise}.\\
\end{array}
\right.
\end{equation}
\begin{equation}
\label{eq:main_carrier_mapping_2}
y_{k}(f_{c} - \beta) = \left\{ \begin{array}{cccccccccc}
y_{k, main}, & \mbox{ if } b_{k} = 0;\\
y_{k, side}, & \mbox{otherwise}.\\
\end{array}
\right.
\end{equation}
In \eqref{eq:main_carrier_mapping_1} and \eqref{eq:main_carrier_mapping_2}, we have assumed that the two tones $f_{c} - \beta$ and $f_{c} + \beta$ experience identical channel realization (assuming a large coherence-bandwidth). 

\subsection{Signal Model for BFSK with Attack}
\label{sec:CA_on_FSK_WA}

With CA, we assume that Eve has the knowledge of $\beta$, and she chooses the waveform $w(t)$ so that some energy is added on the received tone $f_{k} \in \mathcal{T}$ and also on the side-tones $f_{k} + 2\beta$ and $f_{k} -2\beta$. Note that Eve does not know whether $f_{k}$ is $f_{c} + \beta$ or $f_{c} - \beta$, for some $f_{c}$. Therefore, she uniformly divides her energy on both the side-tones $f_{k} + 2\beta$ and $f_{k} - 2\beta$ so that the attack is successful irrespective of the transmitted bit. Assuming  frequency-flat equivalent channel at Bob, the received symbols under the CA are given by

\begin{small}
\begin{eqnarray}
\label{eq:FSK_rx_signal_with_attack}
y_{k, main} & = & \sqrt{E_{Alice}}h^{(AB)}_{k} + \sqrt{\alpha\theta E_{Alice}}h^{(AEB)}_{k, main} + \sqrt{\alpha\theta}n^{(EB)}_{k} \nonumber \\
& & + n^{(B)}_{k, main},\\
y_{k, side} & = & \sqrt{\frac{(1 - \alpha)\theta E_{Alice}}{2}}h^{(AEB)}_{k, side} + \sqrt{\frac{(1 - \alpha)\theta}{2}}n^{(EB)}_{k} \nonumber \\ 
& & + n^{(B)}_{k, side},
\end{eqnarray}
\end{small}

\noindent where $\sqrt{\alpha \theta}$ and $\sqrt{\frac{(1 - \alpha)}{2} \theta}$ are the gains applied by Eve on the received tone and on either of the side-tones, respectively. Based on the nature of operations at Eve, we model the complex channels $h^{(AEB)}_{k, main}$ and $h^{(AEB)}_{k, side}$ as $$h^{(AEB)}_{k, main} \triangleq h^{(AE)}_{k}w_{k}h^{(EB)}_{k} \mbox{ and } h^{(AEB)}_{k, side} \triangleq h^{(AE)}_{k}u_{k}h^{(EB)}_{k},$$
respectively, where $h^{(AE)}_{k}$ is the channel from Alice to Eve, distributed as $\mathcal{CN}(0, 1)$, $h^{(EB)}_{k}$ is the channel from Eve to Bob, distributed as $\mathcal{CN}(0, 1)$, and $w_{k}$ and $u_{k}$ are the statistically independent random variables obtained from the waveform $w(t)$. Similarly, the forwarded AWGN  components from Eve are $n^{(EB)}_{k, main} \triangleq h^{(EB)}_{k}w_{k}n^{(E)}_{k}$ and $n^{(EB)}_{k, side} \triangleq h^{(EB)}_{k}u_{k}n^{(E)}_{k}$, where $n^{(E)}_{k}$ is distributed as $\mathcal{CN}(0, \sigma^{2}_{E})$. Henceforth, we denote $\alpha \theta E_{Alice}$ and $\frac{(1 - \alpha)}{2} \theta E_{Alice}$ as $E_{Eve, main}$ and $E_{Eve, side}$, respectively.

To decode the information bits, Bob uses non-coherent energy detection rule given by
\begin{equation}
\label{eq:bfsk_decoding}
\hat{b}_{k} = \left\{ \begin{array}{cccccccccc}
1, & \mbox{ if } |y_{k}(f_{c} + \beta)|^2 > |y_{k}(f_{c} - \beta)|^2;\\
0,  & \mbox{0therwise}.\\
\end{array},
\right.
\end{equation}
where $\hat{b}_{k}$ denotes the estimate of $b_{k}$, and $y_{k}(f_{c} + \beta)$ and $y_{k}(f_{c} - \beta)$ are as given in \eqref{eq:main_carrier_mapping_1} and \eqref{eq:main_carrier_mapping_2}, respectively. 

\subsection{Events Affecting Error-Performance at Bob}
\label{subsec:cross_1_0_BFSK}

\begin{figure}
\begin{center}
\includegraphics[scale=0.37]{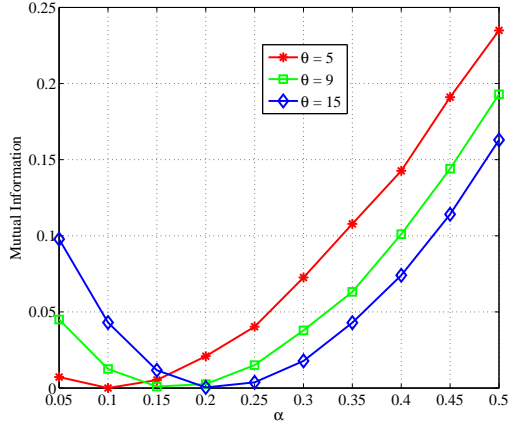}
\vspace{-0.4cm}
\caption{\label{fig:alpha_analysis}Numerically computed values of $\alpha$ that maximize the impact of the attack. The best attack strategy is to pour higher energy on the side channels as the main channel already contains energy contributed by Alice.}
\end{center}
\end{figure}

In the CA, Eve can vary her energy levels $E_{Eve, main}$ and $E_{Eve, side}$ so that the energy levels of the main channel $\sqrt{E_{Alice}}h^{(AB)}_{k} + \sqrt{E_{Eve, main}}h^{(AEB)}_{k, main}$ and the side channel $\sqrt{E_{Eve, side}}h^{(AEB)}_{k, side}$ are close enough to create confusion at Bob. Since Alice and Bob hop across a wide range of narrowbands, Eve cannot learn the narrowband instantaneously, and therefore she cannot force the channel realizations to a specific value with probability one. However, she can change the distribution of the received energy on the tones $f_{c} + \beta$ and $f_{c} - \beta$ at Bob by varying the value of $\alpha$. Along that direction, an interesting question is: \emph{What is the optimal value of $\alpha$ that degrades the error-performance at Bob?}

We now discuss the right choice of the parameter $\alpha$ from the attacker's perspective. With no additive-noise at Eve and Bob, the received energy on the tone chosen by Alice is
\begin{equation}
\label{eq:energy_main_tone}
E_{main} = E_{Alice}\left|h^{(AB)}_{k} + \sqrt{\alpha \theta}h^{(AE)}_{k}w_{k}h_{k}^{(EB)}\right|^{2}.
\end{equation}
Similarly, the received energy on the complementary tone is
\begin{equation}
\label{eq:energy_side_tone}
E_{side} = E_{Alice}\left|\sqrt{\frac{1 - \alpha}{2} \theta}h^{(AE)}_{k}u_{k}h_{k}^{(EB)}\right|^{2}.
\end{equation}
Since $E_{main}$ and $E_{side}$ are random variables, the objective of the attacker is to choose $\alpha$ such that 
\begin{equation*}
p_{cross} \triangleq \mbox{Prob}(E_{main} > E_{side}) = \frac{1}{2}.
\end{equation*}
In other words, the objective is to drive the mutual information $I(b_{k}; \hat{b}_{k}) = 1 - H(p_{cross}) = 0$, where $H(\cdot)$ is the entropy function. We have empirically computed the mutual information values $I(b_{k}; \hat{b}_{k})$ against various values of $\alpha$ over an ensemble of realizations of the random variables $h^{(AB)}_{k}, h^{(AE)}_{k}$, $h^{(EB)}_{k}$, $w_{k}$ and $u_{k}$. To generate the numerical results, we use $E_{Alice} = 1$ and $\theta = 5, 9, 15$. We assume that $w_{k}$ and $u_{k}$ are independent and distributed as $\mathcal{CN}(0, 1)$. The computed mutual information values are presented in Fig. \ref{fig:alpha_analysis} as a function of $\alpha$ for several values of $\theta$. The plots suggest that the attack impact can be maximized by choosing $\alpha = 0.1, 0.15, 0.2$ for $\theta = 5, 9, 15$, respectively. Note that larger value of $\alpha$ decreases the cross-over probability, thereby increasing the mutual information of the channel. Similarly, values of $\alpha$ lower than the above optimal values increases the cross-over probability more than $0.5$, using which the receiver can achieve cross-over probability less than $0.5$ by flipping the decoded bits on each symbol.

Since the density functions on $E_{main}$ and $E_{side}$ are not analytically tractable, the following theorem provides closed-form expressions on sub-optimal values of $\alpha$ by approximating $h^{(AEB)}_{k, main}$ and $h^{(AEB)}_{k, side}$ to be Gaussian distributed. 

\begin{figure*}
\begin{equation}
\label{eq:averaged_transition_probability}
\mathbb{E}_{|h^{(AEB)}_{side}|^{2}} \left(\mbox{Prob}\left(E_{main} > \left(\frac{1 - \alpha}{2}\right) \theta |h^{(AEB)}_{side}|^{2}\right)\right) = \frac{2 + 2\alpha \theta }{2 + \alpha \theta + \theta}
\end{equation}
\hrule
\end{figure*}

\begin{theorem}
\label{thm_approx}
Under the assumption that $h^{(AEB)}_{k, main}$ and $h^{(AEB)}_{k, side}$ are statistically independent complex Gaussian random variables, with $\sigma^{2}_{Eve} = \sigma^{2}_{Bob} = 0$, the optimal value of $\alpha$ (denoted by $\alpha^{*}$) which minimizes the mutual information $I(b_{k}; \bar{b}_{k})$ is $\frac{\theta - 2}{2\theta}$.
\end{theorem}
\begin{IEEEproof}
In \eqref{eq:energy_main_tone} and \eqref{eq:energy_side_tone}, the variables $w_{k}$ and $u_{k}$ are statistically independent with zero mean and unit variance. We note that the random variables $h^{(AE)}_{k}u_{k}h_{k}^{(EB)}$ and $h^{(AE)}_{k}w_{k}h_{k}^{(EB)}$ are uncorrelated but not statistically independent. Furthermore, since the distribution on the product of the three Gaussian random variables is not tractable, we seek to obtain $\alpha$ by assuming that $h^{(AE)}_{k}u_{k}h_{k}^{(EB)}$ and $h^{(AE)}_{k}w_{k}h_{k}^{(EB)}$ are Gaussian distributed and statistically independent. We need to choose $\alpha$ such that 
$\mbox{Prob}(E_{main} > E_{side}) = 0.5.$ From \eqref{eq:energy_main_tone} and \eqref{eq:energy_side_tone}, we note that $E_{Alice}$ is a common factor in both $E_{main}$ and $E_{side}$, and therefore, without loss of generality, we assume $E_{Alice} = 1$. In the rest of this proof, we compute $\mbox{Prob}(E_{main} > E_{side})$ as a function of $\alpha$ and $\theta$, and subsequently obtain $\alpha$ that achieves cross-over probability $0.5$. First, we compute $\mbox{Prob}(E_{main} > E_{side} = e_{side}),$ where $e_{side}$ is a realization of the random variable $E_{side}$. Since $E_{main}$ is exponential distributed with mean $(1 + \alpha \theta)$, the above probability can be written as
\begin{equation*}
\mbox{Prob}(E_{main} > E_{side} = e_{side}) = e^{\frac{-e_{side}}{1 + \alpha \theta}}.
\end{equation*}
Replacing $e_{side}$ by $\frac{1 - \alpha}{2} \theta |h^{(AEB)}_{side}|^{2}$, we have 
\begin{equation*}
\mbox{Prob}\left(E_{main} > \left(\frac{1 - \alpha}{2}\right) \theta |h^{(AEB)}_{side}|^{2}\right) = e^{\frac{-((\frac{1 - \alpha}{2}) \theta |h^{(AEB)}_{side}|^{2})}{1 + \alpha \theta}}.
\end{equation*}
Finally, since $|h^{(AEB)}_{side}|^{2}$ is also exponentially distributed, we take the average of the above expression with respect to $|h^{(AEB)}_{side}|^{2}$ to obtain \eqref{eq:averaged_transition_probability}.
In order to obtain the above probability to be $0.5$, the variable $\alpha^{*}$ must satisfy the constraint $\alpha^{*} = \frac{\theta -2}{2\theta}.$
\end{IEEEproof}


\subsection{Simulation Results}
\label{sec:sims_BFSK}

In this section, we demonstrate the error-performance of BFSK based FH under CA. To carry out the experiments, we assume that the channels $\{h^{(AB)}_{k, j}(f_{c}) ~|~ f_{c} \in \mathcal{F}\}$ across the $N$ narrowbands are statistically independent and distributed as $\mathcal{CN}(0, 1)$. Similarly, $\{h^{(AE)}_{k}(f_{c}) ~|~ f_{c} \in \mathcal{F}\}$ and $\{h^{(EB)}_{k, j}(f_{c}) ~|~ f_{c} \in \mathcal{F}\}$ are also i.i.d. as $\mathcal{CN}(0, 1)$ across the $N$ narrowbands.

\begin{figure}
\begin{center}
\includegraphics[scale=0.4]{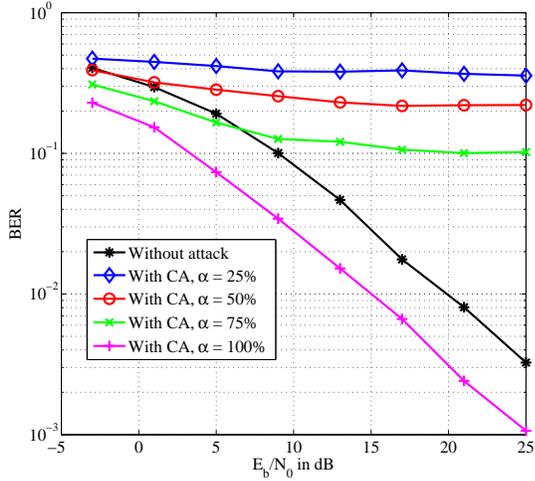}
\vspace{-0.5cm}
\caption{\label{fig:sims_ca} Error-performance of BFSK against CA for various values of $\alpha$. In the CA, $\alpha$ fraction of total energy at Eve is spent on received carrier-frequency $f_{k}$, $\frac{1 - \alpha}{2}$ fraction each on $f_{k} + 2\beta$ and $f_{k} - 2\beta$. Interestingly, $\alpha = 100\%$ outperforms the no-attack case as Eve's signals help Bob in decoding.}
\end{center}
\end{figure}

\begin{figure}
\begin{center}
\includegraphics[scale=0.34]{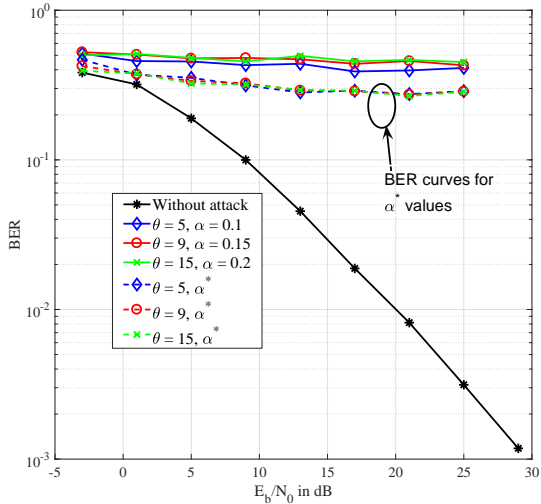}
\vspace{-0.3cm}
\caption{\label{fig:sims_ca_Gaussian_alpha} Error-performance of BFSK against CA when $\alpha$ is computed using Gaussian assumptions on $h^{(AEB)}_{k, main}$ and $h^{(AEB)}_{k, side}$.}
\end{center}
\end{figure}

To showcase the effect of CA, we present the error-performance of: (i) \emph{BFSK based FH without attack} and (ii) \emph{BFSK based FH with CA}. In the latter scheme, Eve executes the CA with various values of $\alpha$. Specifically, $\alpha$ fraction of the total energy at Eve is used for corrupting the main channel, while $\frac{1-\alpha}{2}$ fraction of it is used to introduce additional energy on either of the side-tones. We use $\theta = 9$, i.e., as Alice increases $E_{Alice}$, Eve also increases $E_{Eve}$ proportionately. For the AWGN, we use $\sigma^{2}_{Bob} = 1$ and $\sigma^{2}_{Eve} = 0.01$. In Fig. \ref{fig:sims_ca}, we plot the BER curves of the above two schemes against $\frac{E_{b}}{N_{0}} = \frac{E_{Alice}}{\sigma^{2}_{Bob}}$. The plots confirm our observations from the previous section that it is a better strategy for Eve to use $\alpha$ that ensures comparable energy levels on the two tones $f_{c} + \beta$ and $f_{c} - \beta$. In Fig. \ref{fig:sims_ca_Gaussian_alpha}, we also plot the BER when closed-form expression on $\alpha^{*}$ (from Theorem 1) is used. The plots show that although these values of $\alpha$ are not optimal, they continue to result in degraded error-performance at Bob.

\subsection{Mitigation Strategy: Enhanced BFSK}
\label{sec:mitigation}

In the existing BFSK based FH scheme, randomness is applied on the choice of the carrier-frequency $f_{c} \in \mathcal{F}$ but not on the choice of the tones $f_{c} + \beta$ and $f_{c} - \beta$. This implies, given $f_{k}$ at Eve, the tone on which bit-$\bar{b}_{k}$ is encoded is deterministic upto $1$ bit of randomness, and this weakness is specifically exploited by Eve during the CA. Therefore, from the legitimate users' perspective, they must obfuscate the location of the tones $f_{c} + \beta$ and $f_{c} - \beta$. This way, Eve cannot introduce significant energy on the tone that carries bit-$\bar{b}_{k}$. To achieve this mitigation strategy, we enable Alice and Bob to share a secret-key using which the random positions of the side-tones are determined. As a consequence, Eve cannot learn their locations. Meanwhile, Bob's strategy is to observe the appropriate pair of tones, and then apply non-coherent energy detection to decode the information bits.

\begin{figure}[h]
\begin{center}
\includegraphics[scale=0.37]{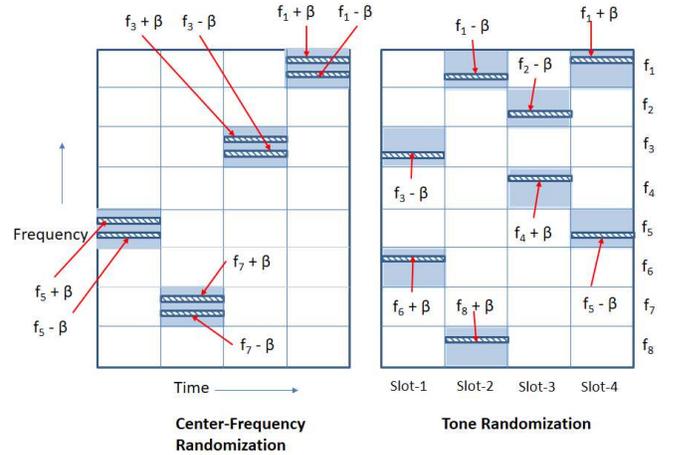}
\vspace{-0.6cm}
\caption{\label{fig:EFH}Pictorial representation of traditional BFSK (left-side) and enhanced BFSK (right-side) with $8$ carrier-frequencies and $4$ time-slots. With EBFSK, tone randomization helps in mitigating the CA on BFSK based FH.}
\end{center}
\end{figure}

We refer to the proposed mitigation strategy as Enhanced BFSK (EBFSK), wherein a pair of tones is randomly chosen from the set $\mathcal{T}$ based on a secret-key, and one of them is used to communicate bit-$1$ and the other for bit-$0$. Since this selection is based on a shared-key, Bob observes the symbols on the chosen tones, and then decodes the information based on the received energy. To illustrate the above idea, we use the example depicted in Fig. \ref{fig:EFH}. As shown on the left-hand side of Fig. \ref{fig:EFH}, traditional BFSK based FH randomizes only the carrier-frequencies, while keeping the side-tones fixed. On the other hand, the EBFSK scheme randomizes the side-tones as shown on the right-hand side of Fig. \ref{fig:EFH}. For instance, in time-slot $4$, bit-$1$ can be encoded on the tone $f_{5} - \beta$ and bit-$0$ can be encoded on the tone $f_{1} + \beta$. With that mapping, when the attacker instantaneously receives the signal on $f_{5} - \beta$, she does not know the location of the other tone which carries bit-$0$. As a result, executing the CA given in Section \ref{sec:CA_on_FSK} does not guarantee degradation of error-performance at Bob.

\begin{figure}[h]
\begin{center}
\includegraphics[scale=0.31]{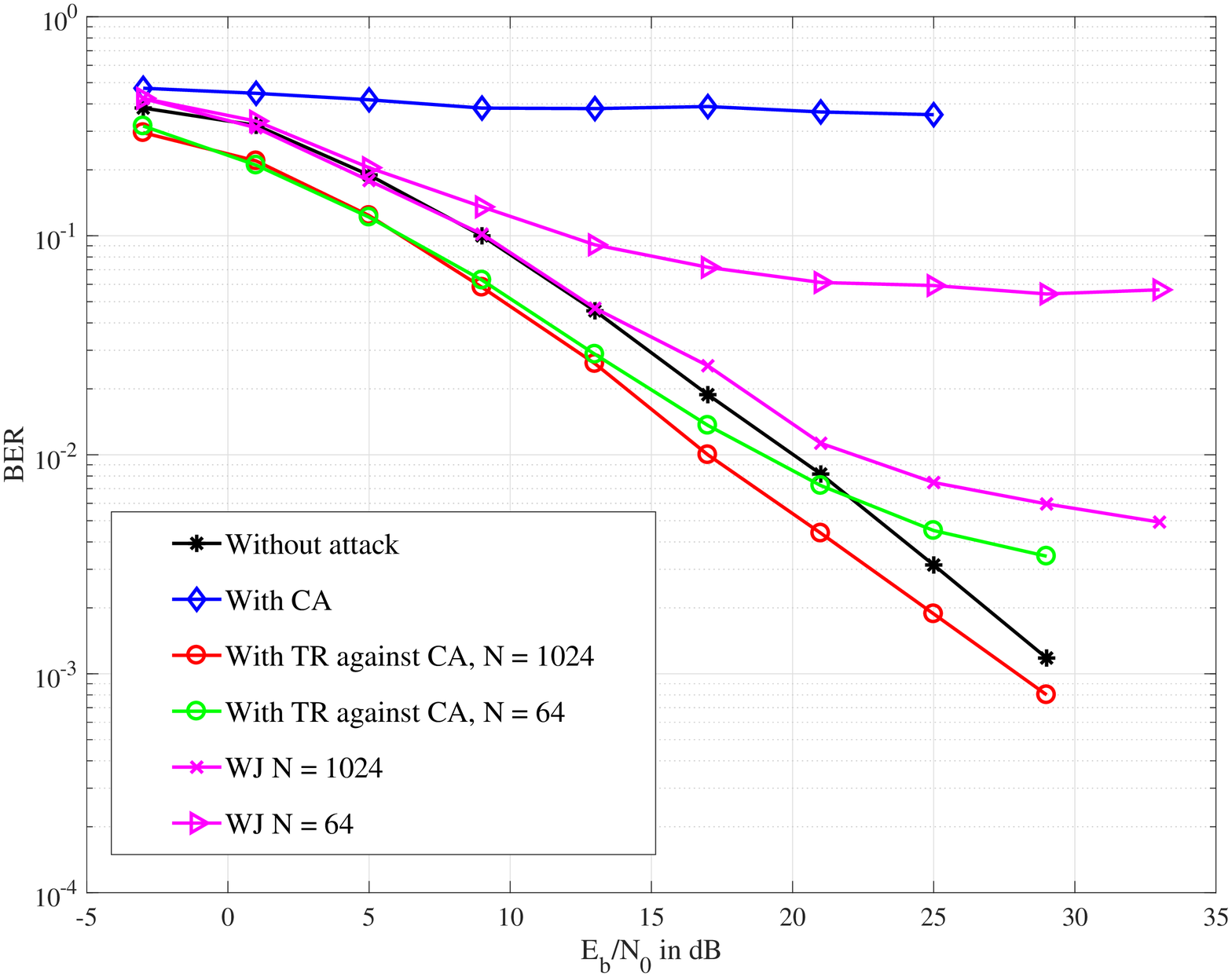}
\vspace{-0.9cm}
\caption{\label{fig:ber_EFH_CA} Error-performance of Enhanced BFSK with tone randomization (TR) against CA with $\alpha = 0.25$. With EBFSK, tone randomization helps in mitigating the CA on FH.}
\end{center}
\end{figure}

To demonstrate the impact of tone randomization, we present the error-performance of: (i) \emph{CA on traditional FH based BFSK}: BFSK based FH is employed at Alice, wherein only the carrier-frequencies are subject to randomization. (ii) \emph{CA on Enhanced BFSK}: BFSK based FH is employed at Alice, wherein all the $2N$ tones of $\mathcal{T}$ are subject to randomization. For the above two schemes, CA is executed as described in Section \ref{sec:CA_on_FSK} with $\alpha = 25\%$. In Fig. \ref{fig:ber_EFH_CA}, we plot the uncoded BER curves of the above schemes with $N = 64$ and $N = 1024$. The plots show that tone randomization assists the legitimate users in mitigating the CA. They also reinforce the point that larger value of $N$ helps in mitigating the CA. This is because, given the tone for bit-$b_{k}$, the probability that either $f_{k} + 2\beta$ or $f_{k} - 2\beta$ is used for bit-$\bar{b}_{k}$ is $\frac{1}{2N-1}$. In contrast, the error-performance of traditional FH under CA does not depend on $N$ as the tones carrying bit-$\bar{b}_{k}$ is deterministic upto $1$ bit randomness.

We now discuss a possible counter-strategy by Eve to overcome the idea of tone randomization. In this strategy, Eve distributes all her energy $E_{Eve} = \theta E_{Alice}$ on wideband jamming. In Fig. \ref{fig:ber_EFH_CA}, we also present the BER performance of EBFSK against wideband jamming when $\theta = 9$ for $N = 64$ and $N = 1024$. The plots show that the error-performance degrades as $E_{b}/N_{o}$ increases, and the degree of degradation depends on the value of $N$. Similar to OOK signalling scheme, BER of BFSK experiences error-floor behaviour with increased jamming energy; this is because equal energy is injected on tones carrying bit-$0$ and bit-$1$.


\begin{table*}
\begin{scriptsize}
\caption{convolution attack on OOK and BFSK}
\vspace{-0.2cm}
\begin{center}
\begin{tabular}{|c|c|c|c|c|c|c|c|c|c|c|}
\hline \textbf{Features} & \textbf{On-Off Keying} & \textbf{Binary Frequency Shift Keying}\\
\hline Attack objective &  Introduce deep fades & Introduce comparable energy levels on tones carrying bit-$0$ and bit-$1$\\
\hline Defense & Use large number of receive antennas at Bob & Use frequency-hopping with tone randomization\\
\hline Shared keys & Needed for frequency hopping & Needed for tone randomization in addition to frequency-hopping\\
\hline Wideband Jamming & Error-floor behaviour at high SNR & Error-floor behaviour at high SNR\\
\hline Generalization to higher-order modulation & Not effective & Fffective\\
\hline Measurement of energy distribution at Bob & Needed to design $E^{*}_{th}$ & Not needed \\
\hline Pilot contamination by Eve & Degrades the performance & Resilient to pilot contamination \\

\hline
\end{tabular}
\end{center}
\label{table:diff_table}
\end{scriptsize}
\end{table*}

\begin{figure}
\begin{center}
\includegraphics[scale=0.30]{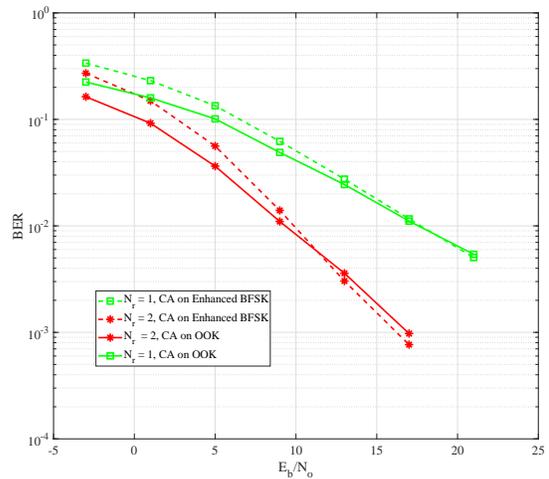}
\vspace{-0.5cm}
\caption{\label{ber_ook_bfsk}Error-performance of OOK and BFSK against corresponding variants of CA on an FH system with $N = 1024$.}
\end{center}
\end{figure}

\section{Discussion}
\label{sec:summary}

We have proposed convolution attacks (CA) on FH based wireless communication by full-duplex radios. We have shown that the CA can convert a slow-fading channel between the transmitter and the receiver to a rapid fading one, thereby forbidding the users to apply amplitude-modulation based signalling schemes. Subsequently, we have studied (i) FH-based OOK and (ii) FH-based BFSK, as countermeasures to mitigate the CA. 

In this concluding section, we discuss some important advantages and disadvantages of the proposed countermeasures, as listed in Table \ref{table:diff_table}. First, we compare the error-performance of the proposed countermeasures against the corresponding variants of CA when $N = 1024$. We have used $N_{r} = 1, 2$ at Bob, and $N_{e} = 1$ at Eve. For the OOK scheme, Eve executes the CA with $\alpha = 1$ and $\theta = 9$, whereas Bob employs non-coherent energy detection to recover the bits. The threshold values for energy detection are computed using the measured energy distributions during the ON and the OFF states. For the BFSK scheme, Eve uses $\alpha = 0.15$ and $\theta = 9$ so that Bob witnesses comparable energy levels on both the tones. As a countermeasure, Alice and Bob employ EBFSK wherein the tones carrying bit-$0$ and bit-$1$ are randomized. When comparing the two schemes, we note that BFSK has lower spectral-efficiency than OOK. To keep the comparison fair, the average transmit energy of the OOK scheme is increased so as to keep the energy per bit constant between the two schemes. We have presented the BER performance of the two schemes in Fig. \ref{ber_ook_bfsk}, which shows that the OOK scheme marginally outperforms BFSK at lower values of $E_{b}/N_{o}$ under the attack, whereas the performance of the two schemes are approximately same when $E_{b}/N_{o}$ is high. In a nutshell, both OOK and BFSK experience similar error-performance against the CA with the exception that (i) BFSK has higher-overhead than OOK as it needs additional shared-key to randomize the tones carrying bit-$1$ and bit-$0$, and (ii) OOK requires Bob to measure the energy distributions to determine the optimal threshold $E^{*}_{th}$, whereas BFSK does not. As a result, if the CRFH attacker does not attack the pilot symbols, then BFSK continues to be effective, whereas OOK may fail due to mismatch between the estimated threshold value and the true energy distributions during the attack.


When generalizing OOK to higher-order modulations, we believe that $M$-PAM (Pulse Amplitude Modulation) variant of OOK for $M > 2$, will result in degraded error-performance against CA. This is because, while symbol $0$ can be detected at Bob, distinguishing non-zero symbols will be a challenge as fading is completely controlled by Eve. However, when generalizing BFSK, we believe that both the attack as well as the countermeasure can be generalized to higher-order FSK. 

\subsection{Directions for Future Research}

For future work, we are interested in studying the impact of convolution attack when the full-duplex radio at Eve experiences imperfect self-interference cancellation. With this constraint, we envisage memory property being introduced by the attacker since the forwarded symbol at a given time-instant will affect the subsequent symbols due to the leakage effect. 

\section*{Acknowledgements}

This work was supported by the Indigenous 5G Test Bed project from the Department of Telecommunications, Ministry of Communications, India.

\end{document}